\documentclass{lmcs}
\pdfoutput=1

\usepackage{lastpage}
\lmcsdoi{16}{2}{13}
\lmcsheading{}{\pageref{LastPage}}{}{}%
{Nov.~15,~2018}{Jun.~23,~2020}{}


\usepackage[utf8]{inputenc}
\usepackage[
    colorlinks,
    linkcolor={red!50!black},
    citecolor={blue!50!black},
    urlcolor={blue!80!black}
]{hyperref}
\usepackage{mathtools}
\usepackage{bm}
\usepackage{prooftree}
\usepackage[linesnumbered, noend, noline, algoruled, algosection]{algorithm2e}
\usepackage{tikz}
\usetikzlibrary{petri,positioning}
\usepackage{enumitem}

\usepackage{macros}
\SetArgSty{textup} 

\begin{document}

\title[Forward Analysis for WSTS, Part III:\texorpdfstring{\@}{} Karp-Miller Trees]{Forward Analysis for WSTS, Part III:\texorpdfstring{\@}{} Karp-Miller Trees\rsuper{*}}
\titlecomment{{\lsuper*}Extended and expanded version of ``Analysis
  for WSTS, Part III:\texorpdfstring{\@}{} Karp-Miller Trees'' by M.~Blondin, A.~Finkel and
  J.~Goubault-Larrecq, in Proc.\ $37^\text{th}$ IARCS Annual
  Conference on Foundations of Software Technology and Theoretical
  Computer Science ({FSTTCS}), 2017.}

\author[M.~Blondin]{Michael Blondin\rsuper{a}}
\author[A.~Finkel]{Alain Finkel\rsuper{{b,c}}}
\author[J.~Goubault-Larrecq]{Jean Goubault-Larrecq\rsuper{b}}

\address{\lsuper{a}Université de Sherbrooke, Canada}
\email{michael.blondin@usherbrooke.ca}
\thanks{M. Blondin was supported by the Fonds de recherche du Quebec –
  Nature et technologies (FRQNT) and by a Discovery Grant from the
  Natural Sciences and Engineering Research Council of Canada (NSERC)}

\address{\lsuper{b}Université Paris-Saclay, ENS Paris-Saclay, CNRS, Laboratoire Spécification et Vérification, France}
\email{\{finkel,goubault\}@lsv.fr}

\address{\lsuper{c}Institut Universitaire de France}

\begin{abstract}
  \noindent This paper is a sequel of ``Forward Analysis for WSTS,
  Part I:\texorpdfstring{\@}{} Completions'' [STACS 2009, LZI Intl.\ Proc.\ in Informatics
    3, 433–444] and ``Forward Analysis for WSTS, Part II:\texorpdfstring{\@}{} Complete
  WSTS'' [Logical Methods in Computer Science 8(3), 2012]. In these
  two papers, we provided a framework to conduct forward reachability
  analyses of WSTS, using finite representations of downward-closed
  sets.  We further develop this framework to obtain a generic
  Karp-Miller algorithm for the new class of very-WSTS\@. This allows us
  to show that coverability sets of very-WSTS can be computed as their
  finite ideal decompositions. Under natural effectiveness
  assumptions, we also show that LTL model checking for very-WSTS is
  decidable. The termination of our procedure rests on a new notion of
  acceleration levels, which we study. We characterize those domains
  that allow for only finitely many accelerations, based on ordinal
  ranks.
\end{abstract}

\maketitle

\section{Introduction}
\subsection{Context}

A well-structured transition system (WSTS) is an infinite
well-quasi-ordered set of states equipped with transition relations
satisfying one of various possible monotonicity properties. WSTS were
introduced in~\cite{F87} for the purpose of capturing properties
common to a wide range of formal models used in verification. Since
their inception, much of the work on WSTS has been dedicated to
identifying generic classes of WSTS for which verification problems
are decidable. Such problems include termination,
boundedness~\cite{F87,F90,FS01} and
coverability~\cite{ACJT96,ACJT00,BFM17,BFM18}. In general, verifying
safety and liveness properties corresponds respectively to deciding
the coverability and the repeated control-state reachability
problems. Coverability can be decided for WSTS by two different
algorithms: the backward algorithm~\cite{ACJT96,ACJT00} and by
combining two forward semi-procedures, one of which enumerates all
downward-closed
invariants~\cite{DBLP:journals/jcss/GeeraertsRB06,BFM17,BFM18}. Repeated
control-state reachability is undecidable for general WSTS, but
decidable for Petri nets by use of the Karp-Miller coverability
tree~\cite{KM67} and the detection of increasing sequences. That
technique fails on well-structured extensions of Petri nets:
generating the Karp-Miller tree does not always terminate on
$\nu$-Petri nets~\cite{Rosa-VelardoMF11}, on reset Petri
nets~\cite{DBLP:conf/icalp/DufourdFS98}, on transfer Petri nets, on
broadcast protocols, and on the depth-bounded
$\pi$-calculus~\cite{HMM14, Rosa-VelardoM12,
  DBLP:conf/vmcai/ZuffereyWH12} which can simulate reset Petri
nets. This is perhaps why little research has been conducted on
coverability tree algorithms and model checking of liveness properties
for general WSTS\@. Nonetheless, some recent Petri nets extensions, \eg
$\omega$-Petri nets~\cite{GHPR15} and unordered data Petri
nets~\cite{HLLLST16}, benefit from algorithms in the style of Karp and
Miller. Hence, there is hope of finding a general framework of WSTS
with Karp-Miller-like algorithms.

\subsection{The Karp-Miller coverability procedure}

In 1967, Karp and Miller~\cite{KM67} proposed what is now known as the
Karp-Miller coverability tree algorithm, which computes a finite
representation (the \emph{clover}) of the downward closure (the
\emph{cover}) of the reachability set of a Petri net. In 1978, Valk
extended the Karp-Miller algorithm to post-self-modifiying nets~\cite{DBLP:conf/icalp/Valk78}, a strict extension of Petri nets. In
1987, the second author proposed a generalization of the Karp-Miller
algorithm that applies to a class of finitely branching WSTS with
strong-strict monotonicity, and having a WSTS completion in which
least upper bounds replace the original Petri nets
$\omega$-accelerations~\cite{F87,F90}. In 2004, Finkel, McKenzie and
Picaronny~\cite{FMP-wstsPN-icomp} applied the framework of~\cite{F90}
to the construction of Karp-Miller trees for strongly increasing
$\omega$-recursive nets, a class generalizing post-self-modifiying
nets. In 2005, Verma and the third author~\cite{VGL-dmtcs05} showed
that the construction of Karp-Miller trees can be extended to
branching vector addition systems with states. In 2009, the second and
the third authors~\cite{FGL12} proposed a non-terminating procedure
that computes the clover of \emph{any} complete WSTS;\@ this procedure
terminates exactly on so-called cover-flattable systems.  Recently,
this framework has been used for defining computable accelerations in
non-terminating Karp-Miller algorithms for both the depth-bounded
$\pi$-calculus~\cite{HMM14} and for $\nu$-Petri nets; terminating
Karp-Miller trees are obtained for strict subclasses.

\subsection{Model checking WSTS}

In 1994, Esparza~\cite{Esp94} showed that model checking the linear
time $\mu$-calculus is decidable for Petri nets by using both the
Karp-Miller algorithm and a decidability result due to Valk and
Jantzen~\cite{VJ85} on infinite $T$-continual sequences in Petri
nets. LTL is undecidable for Petri net extensions such as lossy
channel systems~\cite{DBLP:conf/icalp/AbdullaJ94} and lossy counter
machines~\cite{DBLP:conf/rp/Schnoebelen10}. In 1998, Emerson and
Namjoshi~\cite{EN98} studied the model checking of liveness properties
for complete WSTS, but their procedure is not guaranteed to terminate.
In 2004, Kouzmin, Shilov and Sokolov~\cite{DBLP:conf/time/KouzminSS04}
gave a generic computability result for a fragment of the
$\mu$-calculus; in 2006 and 2013, Bertrand and Schnoebelen~\cite{DBLP:conf/lpar/BaierBS06,DBLP:journals/fmsd/BertrandS13} studied
fixed points in well-structured regular model checking; both~\cite{DBLP:conf/time/KouzminSS04} and~\cite{DBLP:journals/fmsd/BertrandS13} are concerned with formulas with
upward-closed atomic propositions, and do not subsume LTL\@. In 2011,
Chambart, Finkel and Schmitz~\cite{CFS-atpn2011,DBLP:journals/tcs/ChambartFS16} showed that LTL is
decidable for the recursive class of trace-bounded complete WSTS;\@ a
class which does not contain all Petri nets.

\subsection{Our contributions}
\begin{itemize}
\item We define \emph{very-well-structured transition systems
  (very-WSTS)}; a class defined in terms of WSTS completions, and
  which encompasses models such as Petri nets, $\omega$-Petri nets,
  post-self-modifying nets and strongly increasing $\omega$-recursive
  nets. We show that coverability sets of very-WSTS are computable as
  finite sets of ideals.

\item The general clover algorithm of~\cite{FGL12}, based on the ideal
  completion studied in~\cite{FGL-stacs2009}, does not necessarily
  terminate and uses an abstract acceleration enumeration. We give an
  algorithm, the Ideal Karp-Miller algorithm, which organizes
  accelerations within a tree. We show that this algorithm terminates
  under natural order-theoretic and effectiveness conditions, which we
  make explicit. This allows us to unify various versions of
  Karp-Miller algorithms in particular classes of WSTS\@.

\item We identify the crucial notion of \emph{acceleration level} of
  an ideal, and relate it to ordinal ranks of sets of reachable states
  in the completion. We show, notably, that termination is equivalent
  to the rank being strictly smaller than $\omega^2$. This classifies
  WSTS into those with high rank (the bad ones), among which those
  whose sets of states consist of words (\eg, lossy channel systems)
  or multisets; and those with low rank (the good ones), among which
  Petri nets and post-self-modifying nets.

\item We show that the downward closure of the trace language of a
  very-WSTS is computable, again as a finite union of ideals. This
  shows that downward trace inclusion is decidable for very-WSTS\@.

\item Finally, we prove the decidability of model checking liveness
  properties for very-WSTS under some effectiveness hypotheses.
\end{itemize}

\subsection{A short story of well-structured transition systems}

\emph{Structured transition systems} were initially defined and
studied in~\cite{Fin86,F87,F90} as monotone transition systems
equipped with a well-quasi-ordering on their set of
states. Termination was shown decidable for \emph{structured
  transition systems} with \emph{transitive} monotonicity, while
boundedness was shown decidable for structured transition systems with
\emph{strict} monotonicity. For a subclass of finitely branching
labeled structured transition systems with strong-strict monotonicity,
initially called \emph{well structured transition systems} in~\cite{Fin86,F87,F90}, a generalization of the Karp-Miller algorithm
was shown to compute their coverability sets. In~\cite{ACJT96,ACJT00},
the coverability problem was shown to be decidable for
\emph{well-structured systems}~\cite[Def.~3.1]{ACJT96},
\ie\ \emph{labeled} structured transition systems with \emph{strong
  monotonicity} and satisfying an additional \emph{effective
  hypothesis}: the existence of an algorithm to compute the finite set
of minimal elements $\min(\pred{\upc{s}})$, where $\pred{\upc{s}}$ is
the set of immediate predecessors of the upward-closure $\upc{s}$ of a
state $s$.
 In~\cite{FS01}, mathematical properties were
distinguished from effective properties, and the coverability problem
was shown decidable for the \emph{entire} class of structured
transition systems satisfying the additional \emph{effective
  hypothesis} that there exists an algorithm to compute the finite set
$\min(\upc{\pred{\upc{s}}})$, \ie\ the hypotheses of transitions
labeling and strong monotonicity made in~\cite{ACJT96} turned out to
be superfluous.

Today, following the presentation of~\cite{FS01}, what is
\emph{mathematically} known as \emph{well structured transition
  systems} is exactly the original class of \emph{structured
  transition systems}; and necessary effective hypotheses are added
for obtaining decidability of properties such as termination,
coverability and boundedness.

\subsection{Differences between very-WSTS and WSTS of\texorpdfstring{~\cite{F90}}{ (Finkel, 1990)}}

The class of WSTS of~\cite[Def.~4.17]{F90} is reminiscent of
very-WSTS\@. It requires WSTS to be finitely branching and strictly
monotone, whereas our definition allows infinite branching and
requires the
\emph{completion} to be strictly monotone. Moreover,~\cite[Thm.~4.18]{F90}, which claims that its Karp-Miller procedure
terminates, is incorrect since it does not terminate on transfer Petri
nets and broadcast protocols~\cite{esparza99}, which are finitely
branching and strictly monotone WSTS\@. Finally, some assumptions
required to make the Karp-Miller procedure of~\cite{F90} effective are
missing.


\section{Preliminaries}
We write $\subseteq$ for set inclusion and $\subset$ for strict set
inclusion.  A relation $\leq\ \subseteq X \times X$ over a set $X$ is
a \emph{quasi-ordering} if it is reflexive and transitive, and a
\emph{partial ordering} if it is antisymmetric as well.  It is
\emph{well-founded} if it has no infinite descending chain.  A
quasi-ordering $\leq$ is a \emph{well-quasi-ordering} (resp.\
\emph{well partial order}), \emph{wqo} (resp.\ \emph{wpo}) for short,
if for every infinite sequence $x_0, x_1, \dots \in X$, there exist
$i < j$ such that $x_i \leq x_j$.  This is strictly stronger than
being well-founded.

One example of well-quasi-ordering is the componentwise ordering of
tuples over $\N$. More formally, $\N^d$ is well-quasi-ordered by
$\leq$ where, for every $\vec{x}, \vec{y} \in \N^d$,
$\vec{x} \leq \vec{y}$ if and only if $\vec{x}(i) \leq \vec{y}(i)$ for
every $i \in [d]$. We extend $\N$ to
$\Nomega \defeq \N \cup \{\omega\}$ where $n \leq \omega$ for every
$n \in \Nomega$. $\Nomega^d$ ordered componentwise is also
well-quasi-ordered. Let $\Sigma$ be a finite alphabet. We write
$\Sigma^*$, $\Sigma^+$ and $\Sigma^\omega$ to denote the set of finite
words, nonempty finite words and infinite words over $\Sigma$,
respectively. For every (finite or infinite) nonempty word $w$, we
write $w_i$ to denote its $i^\text{th}$ letter. For every $u,
v \in \Sigma^*$, we write $u \preceq v$ if $u$ is a subword of
$v$, \ie\ $u$ can be obtained from $v$ by removing zero, one or
multiple letters. $\Sigma^*$ is well-quasi-ordered by $\preceq$.

\subsection{Transition systems}

A \emph{(labeled) transition system} is a triple $\S = (X,
\trans{\Sigma})$ such that $X$ is a set, $\Sigma$ is a finite
alphabet, and $\trans{a}\ \subseteq X \times X$ for every $a \in
\Sigma$. Elements of $X$ are called the \emph{states} of $\S$, and
each $\trans{a}$ is a \emph{transition relation} of $\S$. A
\emph{class $\C$ of transition systems} is any set of transition
systems. We extend transition relations to sequences over $\Sigma$,
\ie for every $x, y \in X$, $x \trans{\varepsilon} x$, and $x
\trans{wa} y$ if there exists $x' \in X$ such that $x \trans{w} x'
\trans{a} y$. We write $x \trans{*} y$ (resp.\ $x \trans{+} y$) if
there exists $w \in \Sigma^*$ (resp.\ $w \in \Sigma^+$) such that $x
\trans{w} y$. The finite and infinite \emph{traces} of a transition
system $\S$ from a state $x \in X$ are respectively defined as
\begin{align*}
  \traces{\S}{x} &\defeq \{w \in \Sigma^* : x \trans{w} y \text{ for
    some } y \in X\}, \text{ and} \\
  \tracesinf{\S}{x} &\defeq \{w \in \Sigma^\omega : x \trans{w_1} x_1
  \trans{w_2} \cdots \text{ for some } x_1, x_2, \ldots \in X\}.
\end{align*}
We define the \emph{immediate successors} and \emph{immediate
  predecessors} of a state $x$ under some sequence $w \in \Sigma^*$ as
\begin{align*}
  \ssucc{x}[w] &\defeq \{y \in X : x \trans{w} y\}, \text{ and} \\
  \spred{x}[w] &\defeq \{y \in X : y \trans{w} x\}.
\end{align*}
The \emph{successors} and \emph{predecessors} of $x \in X$ are
\begin{align*}
  \ssucc[*]{x} &\defeq \{y \in X : x \trans{*} y\}, \text{ and} \\
  \spred[*]{x} &\defeq \{y \in X : y \trans{*} x\}.
\end{align*}
These notations are naturally extended to sets, \eg $\ssucc{A}[w]
\defeq \bigcup_{x \in A} \ssucc{x}[w]$.

We say that $\S$ is \emph{deterministic} if $|\ssucc{x}[a]| \leq 1$
for every $x \in X$ and $a \in \Sigma$. When $\S$ is deterministic,
each $a \in \Sigma$ induces a partial function $t_a : X \to X$ such
that $t_a(x) = y$ for each $x \in X$ such that $\ssucc{x}[a]
= \{y\}$. For readability, we simply write $a$ for $t_a$, \ie $a(x) =
t_a(x)$. For every $w \in
\Sigma^*$, we write $w(x)$ for $\ssucc{x}[w]$ if $\ssucc{x}[w] \not=
\emptyset$.

\subsection{Well-structured transition systems}

An \emph{ordered (labeled) transition system} is a triple $(X,
\trans{\Sigma}, \leq)$ such that $(X, \trans{\Sigma})$ is a
\emph{(labeled) transition system} and $\leq$ is a quasi-ordering. An
ordered transition system $\S$ is a \emph{well-structured transition
  system (WSTS)} if $\leq$ is a well-quasi-ordering and $\S$ is
\emph{monotone}, \ie\ for all $x, x', y \in X$ and $a \in \Sigma$ such
that $x \trans{a} y$ and $x' \geq x$, there exists $y' \in X$ such
that $x' \trans{*} y'$ and $y' \geq y$. Many other types of
monotonicities were defined in the literature (see \eg~\cite{FS01}),
but, for our purposes, we only need to introduce strong
monotonicities. We say that $\S$ has \emph{strong monotonicity} if for
all $x, x', y \in X$ and $a \in \Sigma$, $x \trans{a} y$ and $x' \geq
x$ implies $x' \trans{a} y'$ for some $y' \geq y$. We say that $\S$
has \emph{strong-strict monotonicity}\footnote{Strong-strict
  monotonicity should not be confused with strong \emph{and} strict
  monotonicities. Here strongness and strictness have to hold at the
  \emph{same} time.} if it has strong monotonicity and for all $x, x',
y \in X$ and $a \in \Sigma$, $x \trans{a} y$ and $x' > x$ implies $x'
\trans{a} y'$ for some $y' > y$.

\paragraph*{\emph{Remark}.} Although the coverability problem is decidable for \emph{unlabeled} WSTS, we consider labeled WSTS here for two main reasons: firstly, we study the traces of WSTS and their model checking, hence their transitions must be labeled with a finite alphabet; secondly, we extend the acceleration technique to compute the downward closure of reachability sets: we need a labeling of transitions to properly define the acceleration of a sequence of transitions (this labeling is not necessary for Petri nets, but in an abstract model like WSTS, the labeling seems necessary).

\subsection{Verification problems}

We say that a \emph{target state} $y \in X$ is \emph{coverable from}
an \emph{initial state} $x \in X$ if there exists $z \geq y$ such that
$x \trans{*} z$. The \emph{coverability problem} asks whether a target
state $y$ is coverable from an initial state $x$. The \emph{repeated
  coverability problem} asks whether a target state $y$ is coverable
infinitely often from an initial state $x$; \ie whether there exist
$z_0, z_1, \dots \in X$ such that $x \trans{*} z_0 \trans{+} z_1
\trans{+} \cdots$ and $z_i \geq y$ for every $i \in \N$.


\section{An investigation of the Karp-Miller algorithm}
In order to present our Karp-Miller algorithm for WSTS, we first
highlight the key components of the Karp-Miller algorithm for Petri
nets. A \emph{Petri net} with $d$ places is a WSTS $\V = (\N^d,
\trans{T}, \leq)$ induced by a finite set $T \subseteq \N^d \times
\N^d$ and the rules:
\begin{align*}
  \vec{x} \trans{t} \vec{y} \defiff \vec{x} \geq \vec{pre} \land \vec{y}
  = \vec{x} - \vec{pre} + \vec{post} && \text{for every
  } \vec{x}, \vec{y} \in \N^d, t = (\vec{pre}, \vec{post}) \in T.
\end{align*}
Petri nets are deterministic and have strong-strict
monotonicity. Given a Petri net with $d$ places and a vector
$\vec{x}_\text{init} \in \N^d$, the Karp-Miller algorithm initializes
a rooted tree whose root is labeled by $\vec{x}_\text{init}$. For
every $(\vec{pre}, \vec{post}) \in T$ such that $\vec{x}
\geq \vec{pre}$, a child labeled by $\vec{x} - \vec{pre} + \vec{post}$
is added to the root. This process is repeated successively to the new
nodes. If a newly added node $c: \vec{x}$ has an ancestor $c':
\vec{x}'$ such $\vec{x} = \vec{x}'$, then it is not explored
furthermore. If a newly added node $c: \vec{x}$ has an ancestor $c':
\vec{x}'$ such $\vec{x} > \vec{x}'$, then $c$ is relabeled by the
vector $\vec{y} \in \Nomega^d$ such that $\vec{y}(i) \defeq
\vec{x}(i)$ if $\vec{x}(i) = \vec{x}'(i)$ and $\vec{y}(i) \defeq
\omega$ if $\vec{x}(i) > \vec{x}'(i)$. The latter operation is called
an \emph{acceleration} of $c$ with respect to $c'$.

A vector $\vec{x}_\text{tgt}$ is coverable from $\vec{x}_\text{init}$
if and only if the resulting tree $\T$ contains a node $c: \vec{x}$
such that $\vec{x} \geq \vec{x}_\text{tgt}$. A slightly more complex
characterization in terms of $\T$ further allows to decide whether a
vector $\vec{x}_\text{tgt}$ is repeatedly coverable from
$\vec{x}_\text{init}$.

\subsection{Ideals and completions}\label{ssec:ideals:comp}

One feature of the Karp-Miller algorithm is that it works over
$\Nomega^d$ instead of $\N^d$. Intuitively, vectors containing some
$\omega$ correspond to ``limit'' elements. For a generic WSTS
$\S = (X, \trans{\Sigma},\allowbreak \leq)$, a similar extension of
$X$ is not obvious.  Let us present one, called the \emph{completion}
of $\S$ in~\cite{FGL12}. Instead of operating over $X$, the completion
of $\S$ operates over the so-called \emph{ideals} of $X$. In
particular, the ideals of $\N^d$ are isomorphic to $\Nomega^d$.

Let $X$ be a set quasi-ordered by $\leq$. The \emph{downward closure}
of $D \subseteq X$ is defined as
\begin{align*}
  \downc{D} &\defeq \{x \in X : x \leq y \text{ for some } y \in D\}.
\end{align*}
A subset $D \subseteq X$ is \emph{downward-closed} if $D =
\downc{D}$. An \emph{ideal} is a downward-closed subset $I \subseteq
X$ that is additionally \emph{directed}: $I$ is non-empty and for all
$x, y \in I$, there exists $z \in I$ such that $x \leq z$ and $y \leq
z$ (equivalently, every finite subset of $I$ has an upper bound in
$I$). We denote the set of ideals of $X$ by $\ideals{X}$, \ie
$\ideals{X} \defeq \{D \subseteq X : D = \downc{D} \text{ and } D
\text{ is directed}\}$.

It is known that
\begin{align*}
  \ideals{\N^d} &= \left\{A_1 \times \cdots \times A_d : A_1, \ldots,
  A_d \in \{\downc{n} : n \in \N\} \cup \{\N\}\right\}.
\end{align*}
Therefore, every ideal of $\N^d$ is naturally represented by some
vector of $\Nomega^d$, and vice versa. We write
$\omega\textrm{-rep}(I)$ for this representation, for every $I \in
\ideals{\N^d}$. For example, the ideal $I = \N \times \downc{8} \times
\downc{3} \times \N$ is represented by $\omega\textrm{-rep}(I) =
(\omega, 8, 3, \omega)$.

\smallskip
Downward-closed subsets can often be represented by finitely many
ideals: in fact, the following Theorem~\ref{thm:antichains:ideals} gives a complete characterization of quasi-ordered sets for which every downward closed subset is equal to a finite union of ideals.

\begin{thm}[\cite{ET43,Bo75,Pou79,PZ85,F86,LMP87,BFM17}]\label{thm:antichains:ideals}
  A countable quasi-ordered set $X$ contains no infinite antichain if,
  and only if, every downward closed subset of $X$ is equal to a
  finite union of ideals.
\end{thm}

From this theorem, we immediately deduce a (known) corollary for wqos:

\begin{cor}\label{thm:ideal:decomp}
  Let $X$ be a well-quasi-ordered set. For every downward-closed
  subset $D \subseteq X$, there exist $I_1, I_2, \dots,
  I_n \in \ideals{X}$ such that $D = I_1 \cup I_2 \cup \dots \cup I_n$.
\end{cor}

The existence of such a decomposition has been proved numerous times
(for partial orderings instead of quasi-orderings) in the order theory
community~\cite{Bo75,Pou79,PZ85,F86,LMP87} under different
terminologies, and is a particular case of a more general set theory
result of Erdős and Tarski~\cite{ET43} on the existence of \emph{limit
numbers} between $\aleph_0$ and $2^{\aleph_0}$. The
paper~\cite{DBLP:journals/apal/FrittaionM14} explains in detail the
fact that Theorem~\ref{thm:antichains:ideals} is attributed to Erdős
and Tarski because the difficult direction (left to right) of
Theorem~\ref{thm:antichains:ideals} can be deduced from~\cite[Theorem
1]{ET43}. For the reader interested in a simple and self-contained
proof, we refer to~\cite[Theorem 3.3]{BFM17}. More specifically, this
proof is based on the fact that such decompositions exist in
well-quasi-ordered sets and is reminiscent of Fraïssé's proof
strategy~\cite[Sect.~4.7.2, p.~124]{F86}, which is based
on~\cite[Lemma~2, p.~193]{Bo75}.

Theorem~\ref{thm:antichains:ideals} gives rise to a canonical
decomposition of downward-closed sets. The \emph{ideal decomposition}
of a downward-closed subset $D \subseteq X$ is the set of maximal
ideals contained in $D$ w.r.t.\ inclusion. We denote the ideal
decomposition of $D$ by
$\idealdecomp(D) \defeq \text{max}_{\subseteq}\{I \in \ideals{X} : I
\subseteq D\}$. By Corollary~\ref{thm:ideal:decomp}, $\idealdecomp(D)$
is finite, and $D = \bigcup_{I \in \idealdecomp(D)}
I$. In~\cite{FGL12,BFM18}, the notion of ideal decomposition is used
to define the completion of unlabeled WSTS\@. We slightly extend this
notion to labeled WSTS:\@

\begin{defi}
  Let $\S = (X, \trans{\Sigma}, \leq)$ be a labeled WSTS\@. The
  \emph{completion} of $\S$ is the labeled transition system
  $\comp{\S} = (\ideals{X}, \ctrans{\Sigma}, \subseteq)$ such that
  \begin{align*}
    I \ctrans{a} J \iff J \in \idealdecomp(\downc{\ssucc{I}[a]}).
  \end{align*}
\end{defi}

The completion of a WSTS enjoys numerous properties. In particular, it
has strong monotonicity, and it is finitely
branching~\cite{BFM18}, \ie $\csucc{I}[a]$ is finite for every
$I \in \ideals{X}$ and $a \in \Sigma$. Note that if $\S$ has
strong-strict monotonicity, then this property is not necessarily
preserved by $\comp{\S}$~\cite{BFM18}. Moreover, the completion of a
WSTS may not be a WSTS since $\ideals{X}$ is not always
well-quasi-ordered by $\subseteq$. However, for the vast majority of
models used in formal verification, $\ideals{X}$ is
well-quasi-ordered, and hence completions remain
well-structured. Indeed, $\ideals X$ is well-quasi-ordered if and only
if $X$ is a so-called $\omega^2$-wqo, and widespread wqos, except
possibly graphs under minor embedding, are $\omega^2$-wqo, as
discussed in~\cite{FGL12}. The traces of a WSTS are closely related to
those of its completion:

\begin{prop}[{\cite{BFM18}}]\label{prop:run:comp}
  The following holds for every WSTS $\S = (X, \trans{\Sigma}, \leq)$:
  \begin{enumerate}
    \item For all $x, y \in X$ and $w \in \Sigma^*$, if $x \trans{w}
      y$, then for every ideal $I \supseteq \downc{x}$, there exists
      an ideal $J \supseteq \downc{y}$ such that $I \ctrans{w} J$.

    \item For all $I, J \in \ideals{X}$ and $w \in \Sigma^*$, if $I
      \ctrans{w} J$, then for every $y \in J$, there exist $x \in I,
      y' \in X$ and $w' \in \Sigma^*$ such that $x \trans{w'} y'$ and
      $y' \geq y$. If $\S$ has strong monotonicity, then $w' = w$.

    \item if $\S$ has strong monotonicity, then $\bigcup_{J \in
      \csucc{I, w}} J = \downc{\ssucc{I, w}}$ for all $I \in
      \ideals{X}$ and $w \in \Sigma^*$.

    \item if $\S$ has strong monotonicity, then $\traces{\S}{x} =
      \traces{\comp{\S}}{\downc{x}}$ and $\tracesinf{\S}{x} \subseteq
      \tracesinf{\comp{\S}}{\downc{x}}$ for every $x \in X$.
  \end{enumerate}
\end{prop}

\begin{proof}\leavevmode
  \begin{enumerate}[align=left, widest=110]
  \item[(1--3)] The proofs given in~\cite{BFM18} for unlabeled WSTS
    can be adapted straightforwardly to labeled WSTS\@. For
    completeness, these adaptations are given in the
    appendix.\smallskip

  \item[(4)]
      \begin{itemize}
        \item For every $w \in \traces \S x$, there is a state $y$
          such that $x \trans w y$.  Use (1) on $I = \downc x$: we
          obtain an ideal $J$ such that $I \ctrans w J$, showing that
          $w \in \traces {\comp\S} {\downc x}$.  Conversely, for every
          $w \in \traces {\comp S} {\downc x}$, there is an ideal $J$
          such that $I \ctrans w J$, where $I = \downc x$.  Ideals are
          non-empty, so pick $y \in J$.  By (2), there are states $x'
          \in I$ and $y' \geq y$ such that $x' \trans w y'$.  The fact
          that $x'$ is in $I$, namely that $x' \leq x$, allows us to
          invoke strong monotonicity and obtain a state $y'' \geq y'$
          such that $x \trans w y''$.  In particular, $w$ is in
          $\traces \S x$.

        \item Let $w \in \tracesinf{\Sigma}{x}$. Let $x_0 \defeq x$,
          and let $x_1, x_2, \ldots \in X$ be such that $x \trans{w_1}
          x_1 \trans{w_2} x_2 \trans{w_3} \cdots$. Let $I_0 \defeq
          \downc{x}$. By~(1), there exists an ideal $I_1 \supseteq
          \downc{x_1}$ such that $I_0 \ctrans{w_1} I_1$. This process
          can be repeated using~(1) to obtain $I_{i-1} \ctrans{w_i}
          I_i$ with $I_i \supseteq \downc{x_i}$ for every $i >
          0$. \qedhere
      \end{itemize}
  \end{enumerate}
\end{proof}

\noindent
It is worth noting that if $\S$ is infinitely branching, then an
infinite trace of $\comp{\S}$ from $\downc{x}$ is not necessarily an
infinite trace of $\S$ from $x$ (\eg\ see~\cite{BFM18}).

Whenever the completion of a WSTS $\S$ is deterministic, we will often
write $w(I)$ for $\csucc{I}[w]$ if the latter is nonempty and if there
is no ambiguity with $\ssucc{I, w}$.

\subsection{Levels of ideals}\label{ssec:levels:ideals}

The Karp-Miller algorithm terminates for the following reasons:
$\N_\omega^d$ is well-quasi-ordered and $\omega$'s can only be added
to vectors along a branch at most $d$ times. Loosely speaking, the
latter property means that $\ideals{\N^d}$ has $d+1$ ``levels''. Here,
we generalize this notion. We say that an infinite sequence of ideals
$I_0, I_1, \ldots \in \ideals{X}$ is an \emph{acceleration candidate}
if $I_0 \subset I_1 \subset \cdots$. An acceleration
candidate \emph{is below} $J \in \ideals{X}$ if $I_i \subseteq J$ for
every $i \in \N$, and it \emph{goes through} a set
$A \subseteq \ideals{X}$ if $I_i \in A$ for some $i \in \N$.

\begin{defi}
  The \emph{$n^\text{th}$ level} of $\ideals{X}$ is defined as
  \[
  A_n(\ideals{X}) =
  \begin{cases}
    \emptyset, & n = 0, \\
    \left\{I \in \ideals{X} : \text{every accel.\ candidate
        below $I$ goes through $A_{n-1}$}\right\} & n > 0.
  \end{cases}
  \]
\end{defi}
When $X$ is clear from the context, we will simply write $A_n$ instead
of $A_n(\ideals{X})$. For the specific case of $X = \N^d$, it can be
shown that:
\begin{align*}
  A_1 &= \{I \in \ideals{\N^d} : \emph{$\omega\textrm{-rep}(I)$ has 
    strictly less than $1$ occurrence\phantom{s} of $\omega$}\}, \\
  A_2 &= \{I \in \ideals{\N^d} : \emph{$\omega\textrm{-rep}(I)$ has 
    strictly less than $2$ occurrences of $\omega$}\}, \\
  &\vdotswithin{=}
\end{align*}
Hence, for all $n \geq 0$:
\begin{align*}
  A_n &= \{I \in \ideals{\N^d} : \emph{$\omega\textrm{-rep}(I)$ has 
  strictly  less than $n$ occurrences of $\omega$}\}.
\end{align*}
Therefore, we have $\emptyset \subset A_1 \subset \cdots \subset
A_{d+1} = A_{d+2} = \cdots$ which corresponds to the fact that
$\ideals{\N^d}$ has $d + 1$ different levels. In particular, if we
identify $A_{d+1} $ with $\N_\omega^d$, \ie the set of its
$\omega$-representations, then $A_{d+k}$ is equivalent to
$\N_\omega^d$ for every $k \geq 1$. More formally:

\begin{prop}
  $A_n(\N_\omega^d)$ is the set of $d$-tuples with less than $n$
  components equal to $\omega$.
\end{prop}

\begin{proof}
  Using the fact that $A_n(\N_\omega^d)$ grows as $n$ grows, it
  suffices to show the claim for $n \leq d + 1$. This is shown by
  induction on $n$. The case $n = 0$ is obvious.

  Let $1 \leq n \leq d + 1$. If $\vec{x} \in \N_\omega^d$ has at least
  $n$ components equal to $\omega$, we obtain an acceleration
  candidate by picking an index $j$ such that $\vec{x}(j) = \omega$,
  and forming the tuples $(\vec{x}(1), \ldots, \vec{x}(j-1), i,
  \vec{x}(j+1), \ldots, \vec{x}(d))$ for $i \in \N$. By induction
  hypothesis, these tuples have at least $n - 1$ components equal to
  $\omega$ and therefore cannot be in $A_{n-1} (\N_\omega^d)$. This
  entails that $\vec{x}$ cannot be in $A_n(\N_\omega^d)$.

  Conversely, assume that $\vec{x}$ has less than $n$ components equal
  to $\omega$, say at positions $1, 2, \ldots, k < n$ (the general
  case is obtained by applying a permutation of the indices). There
  are only finitely many tuples $\vec{y} \leq \vec{x}$ that have their
  first $k$ components equal to $\omega$. Therefore any acceleration
  candidate below $\vec{x}$, being infinite, must contain a tuple with
  at most $k - 1$ components equal to $\omega$. Since $k - 1 < n - 1$,
  by induction hypothesis it must go through $A_{n-1}(\N_\omega^d)$,
  showing that $\vec{x} \in A_n(\N_\omega^d)$.
\end{proof}

In general, we observe that ideal levels are monotonic and
downward-closed with respect to ideal inclusion:
\begin{prop}\label{prop:levels:basic}
  The following holds for every $n \in \N$:
  \begin{enumerate}
  \item for every $I, J \in \ideals{X}$, if $I \in A_n$ and $J
    \subseteq I$, then $J \in A_n$,\label{itm:levels:basic:a}

  \item $A_n \subseteq A_{n+1}$.
  \end{enumerate}
\end{prop}

\begin{proof}
  Let $n \in \N$. If $A_n = \emptyset$, then both claims follow
  immediately. Therefore, let us assume that $A_n \not= \emptyset$.

  \begin{enumerate} \item Let $I \in A_n$ and let $J \in \ideals{X}$ be such
  that $J \subseteq I$. We must show that $J \in A_n$. Let $J_0, J_1,
  \ldots$ be an acceleration candidate below $J$. We have $J_i
  \subseteq J \subseteq I$ for every $i \in \N$. Therefore, $J_0, J_1,
  \ldots$ is also below $I$. Since $I \in A_n$, we conclude that $J_0,
  J_1, \ldots$ goes through $A_{n-1}$, and hence that $J \in A_n$.\medskip

  \item Let $I \in A_n$. For the sake of contradiction, suppose $I
    \not\in A_{n+1}$. By assumption, there exists an acceleration
    candidate $I_0, I_1, \ldots$ below $I$ that does not go through
    $A_n$. Note that $I_i \subseteq I$ for every $i \in \N$. By~(1),
    this implies that $I_i \in A_n$ for every $i \in \N$. Therefore,
    we conclude that $I_0, I_1, \ldots$ goes through $A_n$, which is a
    contradiction. \qedhere
  \end{enumerate}
\end{proof}

\noindent
We have seen that $\ideals{\N^d}$ has only $d + 1$ levels,
\ie\ $A_{d+1}(\ideals{\N^d}) = \ideals{\N^d}$. We generalize this
notion as follows:

\begin{defi}
  $\ideals{X}$ has \emph{finitely many levels} if there exists $n \in
  \N$ such that $A_n = \ideals{X}$.
\end{defi}

In the forthcoming sections, we will be interested in sets of ideals
that have finitely many levels. It is however worth mentioning that
there are natural sets $X$ whose ideals do not have finitely many
levels of ideals, even if $\ideals{X}$ is assumed to be countable and
well-quasi-ordered. We postpone this discussion to
Section~\ref{sec:inv:levels} where we will study ideal levels in more
details and in a more abstract setting.

\subsection{Accelerations}\label{ssec:accel}

The last key aspect of the Karp-Miller algorithm is the possibility to
accelerate nodes. In order to generalize this notion, let us briefly
develop some intuition. Recall that a newly added node $c \colon
\vec{x}$ is accelerated if it has an ancestor $c' \colon \vec{x}'$
such that $\vec{x} > \vec{x}'$. Consider the non-empty sequence $w$
labeling the path from $c'$ to $c$. Since Petri nets have strong-strict
monotonicity, both over $\N^d$ and $\Nomega^d$, $w^n(\vec{x})$ is
defined for every $n \in \N$. For example, if $(5, 0, 1) \trans{w} (5,
1, 3)$ is encountered, $(5, 1, 3)$ is replaced by $(5, \omega,
\omega)$. This represents the fact that for every $n \in \N$, there
exists some reachable marking $\vec{y} \geq (5, n, n)$. Note that an
acceleration increases the number of occurrences of $\omega$. In our
example, the ideal $I = \downc{5} \times \downc{1} \times \downc{3}$,
which is of level $0$, is replaced by $I' = \downc{5} \times \N \times
\N$, which is of level $2$. Based on these observations, we extend the
notion of acceleration to completions:

\begin{defi}
  Let $\S = (X, \trans{\Sigma}, \leq)$ be a WSTS such that $\comp{\S}$
  is deterministic, let $I \in \ideals{X}$, and let $w \in \Sigma^+$
  be such that $\csucc{I}[w] \not= \emptyset$. The \emph{acceleration}
  of $I$ under $w$ is defined as:
  \[w^\infty(I) \defeq
  \begin{cases}
     \bigcup_{k \in \N} w^k(I) & \text{if }
     I, w(I), w^2(I), \ldots \text{ is an acceleration candidate}, \\
     I & \text{otherwise.}
  \end{cases}
  \]
\end{defi}

In other words, if $I$ can be accelerated by repeatedly applying $w$,
then its acceleration is the least upper bound of $I \subset w(I)
\subset w^2(I) \subset \cdots$. This least upper bound is also an
ideal:
\begin{prop}
  Let $\S = (X, \trans{\Sigma}, \leq)$ be a WSTS such that $\comp{\S}$
  is deterministic. We have $w^\infty(I) \in \ideals{X}$ for every $I
  \in \ideals{X}$ and $w \in \Sigma^+$ such that $\csucc{I}[w] \not=
  \emptyset$.
\end{prop}

\begin{proof}
  If $w^\infty(I) = I$, then the claim trivially holds. Thus, we may
  assume that $I, w(I), \linebreak w^2(I), \ldots$ is an acceleration
  candidate. Since $w^\infty(I)$ is a union of downward-closed sets,
  it is readily seen to be downward-closed. Let us show that it is
  also directed. Let $x, y \in w^\infty(I)$. There exist
  $k, \ell \in \N$ such that $x \in w^k(I)$ and $y \in
  w^\ell(I)$. Therefore, both $x$ and $y$ are elements of
  $w^{\max(k, \ell)}(I)$. Since $w^{\max(k, \ell)}(I)$ is an ideal,
  there exists $z \in w^{\max(k, \ell)}(I) \subseteq w^\infty(I)$ such
  that $x \leq z$ and $y \leq z$.
\end{proof}

Recall that in the Karp-Miller algorithm for Petri nets, the level of
an ideal remains unchanged when applying a transition, and increases
when accelerated. This holds because the completion of a Petri net has
strong-strict monotonicity. We introduce a more general (\ie weaker)
type of monotonicity that essentially yields the same behaviour.

Let $\S = (X, \trans{\Sigma}, \leq)$ be a WSTS\@. We define
the \emph{level} of an ideal $I \in \ideals{X}$ as follows. If $I \in
A_n$ for some $n \in \N$, then $\lvl{I}$ is the smallest such $n$, and
otherwise $\lvl{I} \defeq \infty$. We say that the completion of $\S$
has \emph{leveled-strong-strict monotonicity} if for every $I, I',
J \in \ideals{X}$ and $w \in \Sigma^*$ such that
$\lvl{I} \neq \infty$, the following holds:
\[
\text{if } I \subset I', I \ctrans{w} J \text { and } \lvl{I}
= \lvl{J}, \text{ then } I' \ctrans{w} J' \text{ for some }
 J' \in \ideals{X} \text{ s.t. } J \subset J'.
\] In other words, leveled-strong-strict monotonicity only requires
 strong-strict monotonicity to hold between ideals of the same
 level.

Petri nets and their completions enjoy strong-strict monotonicity
(hence also leveled-strong-strict monotonicity), but strong-strict
monotonicity is not inherited by the completion of some extensions
such as post-self-modifying nets and $\omega$-Petri nets.

Let us recall the model of post-self-modifying
nets~\cite{DBLP:conf/icalp/Valk78} for which there is a Karp-Miller
algorithm. In post-self-modifying nets, transitions consume tokens as
in Petri nets but they may add the result of applying a (different)
positive affine function in each place. It has been shown
~\cite{FMP-wstsPN-icomp} that post-self-modifying nets are WSTS with
strong-strict monotonicity on $\N^d$. Their completions are still WSTS
with strong monotonicity on $\Nomega^d$, but they are not strictly
monotone on $\Nomega^d$ (contrary to Figure~3 in
~\cite{FMP-wstsPN-icomp}). Let us show here that completions of
post-self-modifying nets are not strictly monotone. Let us consider a
post-self-modifying net $N$ with two places $p_1,p_2$ and an unique
transition $t$ that adds the contents of $p_1$ onto $p_2$. Consider
the two $\omega$-markings $(\omega,0) < (\omega, \omega)$ from
$\Nomega^2$ and the firing of transition $t$, extended on $\Nomega^2$,
from both $\omega$-markings. We obtain $(\omega, 0) \trans{t}
(\omega, \omega)$ and $(\omega, \omega) \trans{t}
(\omega, \omega)$. Since $(\omega, \omega) \not< (\omega, \omega)$,
transition $t$ is not strictly increasing over $\Nomega^2$, even if
$t$ is strictly increasing over $\N^2$. Hence the completion of $N$
does not satisfy strict monotonicity.

Therefore, post-self-modifying nets are WSTS with strong-strict
monotonicity and that their completions are WSTS with strong
monotonicity. However, they are not \emph{strictly} monotone.

\begin{figure}[h]
  \begin{center}
    \begin{tikzpicture}[->, node distance=1.5cm, auto, very thick, scale=1.0, transform shape]
      \node[place, tokens=3] (p) {};

      \node[transition] (t) [below right=0.25cm and 1cm of p] {}
      edge[pre] node[swap] {2} (p)
      ;

      \node[place, tokens=5] (q) [below left=0.25cm and 1cm of t] {}
      edge[post] node[swap] {$\omega$} (t)
      ;

      \node[place, tokens=1] (r) [right=of t] {}
      edge[pre] node[swap] {$\omega$} (t)
      ;

      \node[font=\huge] () [right=0.75cm of r] {$\rightarrow$};
    \end{tikzpicture}
    \hspace{0.75cm}
    \begin{tikzpicture}[->, node distance=1.5cm, auto, very thick, scale=1.0, transform shape]
      \node[place, tokens=1] (p) {};

      \node[transition] (t) [below right=0.25cm and 1cm of p] {}
      edge[pre] node[swap] {2} (p)
      ;

      \node[place, tokens=1] (q) [below left=0.25cm and 1cm of t] {}
      edge[post] node[swap] {$\omega$} (t)
      ;

      \node[place, tokens=4] (r) [right=of t] {}
      edge[pre] node[swap] {$\omega$} (t)
      ;
    \end{tikzpicture}
  \end{center}
  \caption{\emph{Left}: example of an $\omega$-Petri net marked
    (counterclockwise) with $(3, 5, 1)$. \emph{Right}: example of a
    possible marking, \ie $(1, 1, 4)$, obtained after firing the
    unique transition; the other possible markings are $(1, y, z)$
    where $0 \leq y \leq 5$ and $z \geq 1$. Over the completion of the
    same $\omega$-Petri net, the ideal $\downc{3} \times \downc{5}
    \times \downc{1}$ leads to $\downc{1} \times \downc{5} \times \N$
    when firing the unique transition; or equivalently $(3, 5, 1)$
    leads to $(1, 5, \omega)$ in the $\omega$-representation of the
    ideals.}%
  \label{fig:opn}
\end{figure}
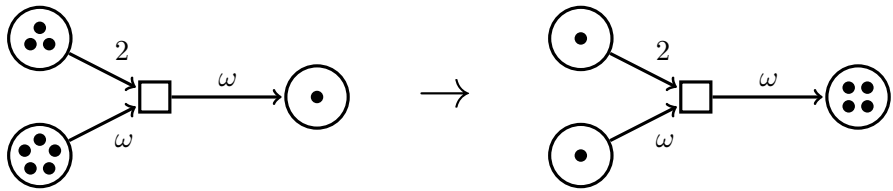

Recall that $\omega$-Petri nets are Petri nets with arcs labeled by
coefficients from $\Nomega$ instead of $\N$. The semantics remains the
same for coefficients over $\N$. Every arc from a place $p$ to a
transition $t$, which is labeled by $\omega$, consumes an arbitrary
number of tokens from $p$ when $t$ is fired. Similarly, every arc from
a transition $t$ to a place $p$, which is labeled by $\omega$,
produces an arbitrary number of tokens in $p$ when $t$ is fired. In
particular, in the completion of an $\omega$-Petri net, an arc from
$t$ to $p$ labeled by $\omega$ increases the contents of $p$ to
$\omega$ whenever $t$ is fired. See Figure~\ref{fig:opn} for an
example of an $\omega$-Petri net, and~\cite{GHPR15} for precise
definitions.

It is known that $\omega$-Petri nets are WSTS with strong-strict
monotonicity and their completions are still WSTS with strong
monotonicity~\cite{BFM18} but they are not \emph{strictly}
monotone. Indeed, consider the $\omega$-Petri net with a single place
$p$ and a unique transition $t$ with a single arc from $t$ to $p$
labeled by $\omega$. We have $\downc{5} \trans{t} \N$, $\downc{6}
\trans{t} \N$, $\downc{5} \subset \downc{6}$, but not $\N \subset
\N$. As a matter of fact, we will prove in Proposition~\ref{post} and
Proposition~\ref{omega} that post-self-modifying nets and
$\omega$-Petri nets have \emph{leveled-strong-strict monotonicity}.

We may now show the following:
\begin{prop}\label{prop:ideals:levels}
  Let $\S = (X, \trans{\Sigma}, \leq)$ be a WSTS such that $\comp{\S}$
  is deterministic and has leveled-strong-strict monotonicity. Let $I
  \in \ideals{X}$ and $w \in \Sigma^+$ be such that $\lvl{I} \not=
  \infty$ and $\csucc{I}[w] \not= \emptyset$. The following
  holds:
  \begin{enumerate}
  \item $\lvl{w(I)} \geq \lvl{I}$,\label{itm:prop:ideals:a}

  \item if $w^\infty(I) \not= I$, then $\lvl{w^\infty(I)} > \lvl{I}$,
  \end{enumerate}
\end{prop}

\begin{proof}\leavevmode
  \begin{enumerate}
  \item We prove the claim by induction on $n = \lvl{I}$. If $n = 1$,
    then the claim trivially holds since $A_0 = \emptyset$ and hence
    $\lvl{J} \geq 1$ for every ideal $J$. Suppose that $n > 1$ and
    that the claim holds for levels smaller than $n$. Since $n > 1$
    and $n$ is the smallest index such that $I \in A_n$, there exists
    an acceleration candidate $I_0, I_1, \ldots$ below $I$ such that
    $I_i \in A_{n - 1}$ and $I_i \not\in A_{n - 2}$ for some $i \in
    \N$. In other words, $\lvl{I_i} = n - 1$.

    Observe that $I_{i+1}, I_{i+2}, \ldots$ is also an acceleration
    candidate below $I$. Thus, there exists $j > i$ such that $I_j \in
    A_{n-1}$. By repeating this process, we obtain an acceleration
    candidate $I_{i_0}, I_{i_1}, \ldots$ below $I$ such that $I_{i_j}
    \in A_{n-1}$ for every $j \in \N$. Thus, by $\lvl{I_0} = n - 1$
    and
    by~Proposition~\ref{prop:levels:basic}~\eqref{itm:levels:basic:a},
    we have $\lvl{I_{i_0}} = \lvl{I_{i_1}} = \cdots = n - 1$. Hence, by
    leveled-strong-strict monotonicity and determinism of $\comp{S}$,
    we have
    \begin{align}
      w(I_{i_0}) \subset w(I_{i_1}) \subset \cdots \subseteq
      w(I).\label{eq:accel:wI}
    \end{align}
    Observe that~\eqref{eq:accel:wI} yields an acceleration candidate
    below $w(I)$. Thus, there exists $\ell \in \N$ such that
    $w(I_{i_\ell}) \in A_{\lvl{w(I)} - 1}$. Therefore, we are done
    since:
    \begin{align*}
      \lvl{w(I)}
      &\geq \lvl{w(I_{i_\ell})} + 1
      && \text{(by $w(I_{i_\ell}) \in A_{\lvl{w(I)} - 1}$)} \\
      &\geq \lvl{I_{i_\ell}} + 1
      && \text{(by ind.\ hyp.\ since $\lvl{I_{i_\ell}} < n$)} \\
      &= (n - 1) + 1 \\
      &= n \\
      &= \lvl{I}.
    \end{align*}

  \item Assume $w^\infty(I) \not= I$. For the sake of contradiction,
    suppose that $\lvl{w^\infty(I)} \leq \lvl{I}$. Since $w^\infty(I)
    \not= I$, the sequence $I, w(I), w^2(I), \ldots$ is an
    acceleration candidate. By definition of $w^\infty(I)$, this
    acceleration candidate is below $w^\infty(I)$. Moreover, it goes
    through $A_{\lvl{I}-1}$, and hence there exists $k \in \N$ such
    that $\lvl{w^k(I)} = \lvl{I} - 1$. This contradicts~(1). \qedhere
  \end{enumerate}
\end{proof}


\section{The Ideal Karp-Miller algorithm}\label{sec:ikm}
We have now introduced all the concepts necessary to present our
generalization of the Karp-Miller algorithm. This algorithm applies to
a new class\footnote{Note that the definition of very-WSTS given here
is slightly more general than the one that appeared in the preliminary
version of this paper~\cite{BFG17}. More precisely, strong-strict
monotonicity is replaced here with leveled-strong-strict monotonicity,
which allows to encompass models such as $\omega$-Petri nets.} of WSTS
that enjoy all of the generalized properties of Petri nets:

\begin{defi}\label{def:very:wsts}
  A \emph{very-WSTS} is a labeled WSTS $\S = (X, \trans{\Sigma},
  \leq)$ such that:
  \begin{itemize}
  \item $\S$ has strong monotonicity,

  \item $\comp{\S}$ is a deterministic WSTS with leveled-strong-strict
    monotonicity,

  \item $\ideals{X}$ has finitely many levels.
  \end{itemize}
\end{defi}

\noindent
Note that the completion $\comp{\S}$ of a WSTS $\S$ always have strong
monotonicity~\cite{BFM18}. However, it does not necessarily have
leveled-strong-strict monotonicity. Moreover, if it does, it is not
necessarily the case that $\S$ has strong monotonicity. In other
words, although the first condition of Definition~\ref{def:very:wsts}
may first appear redundant, it is not the case:

\begin{prop}
  The first condition of Definition~\ref{def:very:wsts} is not
  redundant.
\end{prop}

\begin{proof}
  We construct a WSTS $\S$ that satisfies all the requirements of a
  very-WSTS except for strong monotonicity. Let $\S = (\N,
  \trans{\{t\}}, \leq)$ be the ordered transition system such that $m
  \trans{t} m \div 2$ if $m$ is even, and $m \trans{t} n$ for every $n
  \in \N$ otherwise. Since $\N$ is well-quasi-ordered, it suffices to
  show that $\S$ is monotone in order to show that it is a WSTS\@. Let
  $m, m', n \in \N$ be such that $m \trans{t} n$ and $m' \geq m$. If
  $m'$ is odd, then $m' \trans{t} n$ and we are done. If $m'$ is even,
  then it can be repeatedly halved until some odd number is obtained,
  after which we can reach $n$ in one step, \ie\ $m' \trans{*}
  n$. Observe that $\S$ is not strongly monotone since $1 \trans{t}
  3$, but $3 \not\in \downc{1} = \downc{\ssucc{2}[t]}$.

  Let us now show that $\comp{\S}$ is a deterministic WSTS with
  leveled-strong-strict monotonicity. It is readily seen that
  $\comp{\S}$ is deterministic since: \[\csucc{\downc{0}}[t] =
  \{\downc{0}\} \text{ and } \csucc{I}[t] = \{\N\} \text{ for every
    ideal } I \neq \downc{0}.\] It remains to show that $\comp{\S}$
  has monotonicity and leveled-strong-strict monotonicity. Let $I, I',
  J \in \ideals{\N}$ be such that $I \ctrans{t} J$ and $I \subset
  I'$. Note that $I' \supset I \supseteq \downc{0}$, hence $I'$ must
  contain at least one odd number. Therefore, $I' \ctrans{t} \N$ which
  shows (standard) monotonicity. Since $I \subset I' \subseteq \N$, we
  have $I \neq \N$. Thus, if $\lvl{I} = \lvl{J}$, then we have $J \neq
  \N$. Since $I' \ctrans{t} \N$ and $\N \supset J$, this shows
  leveled-strong-strict monotonicity.
\end{proof}

We claim that the class of very-WSTS includes Petri nets (and hence
vector addition systems with/without states), $\omega$-Petri
nets~\cite{GHPR15}, post-self-modifying
nets~\cite{DBLP:conf/icalp/Valk78} and strongly increasing
$\omega$-recursive nets~\cite{FMP-wstsPN-icomp} for which Karp-Miller
algorithms were known.

Recall that a \emph{strongly increasing function} $f: \N^d \trans{}
\N^d$ is a nondecreasing function defined on an upward closed set of
$\N^d$ that satisfies the following strongly increasing property:
\[ \text{for every}\ \vec{x}, \vec{y} \in \N^d, \text{for every}\ P
\subseteq \{1, 2, \ldots, d\},\ \vec{x} \leq_P \vec{y} \implies
f(\vec{x}) \leq_P f(\vec{y}),
\] where the ordering $\leq_P$ is defined by
\[\vec{x} \leq_P \vec{y} \defiff \vec{x} \leq \vec{y} \land \vec{x}(i)
< \vec{y}(i)\ \text{for every}\ i \in P.
\] A \emph{strongly increasing recursive net} $N$ is a finite set of
strongly increasing recursive functions. A \emph{strongly increasing
  $\omega$-recursive net} $N$ is a strongly increasing recursive net
such that the continuous extensions of the functions $ f: \Nomega^d
\trans{} \Nomega^d$ satisfy the previous strongly increasing property
but over $\Nomega^d$ instead (see, \eg,~\cite{FGL12} for a definition
of continuous extension). Let us write $\S_N$ for the transition
system naturally associated with a net $N$. We may observe that
$\S_{\comp{N}} = \comp{\S}_{\!N}$.

Since nondecreasing functions of post-self-modifying nets and of
strongly increasing $\omega$-recursive nets are incomparable, we
define another class of nondecreasing functions that subsumes the two
previous ones. Let us identify the nondecreasing functions over $\N^d$
that are strictly increasing, but only on the subset of $\N^d$ such
that $\vec{x}$ and $f(\vec{x})$ have the same number of $\omega$'s.

\begin{defi}
  A \emph{leveled-increasing} partial function $f \colon \N^d \to
  \N^d$ is a nondecreasing partial function such that its continuous
  extension $\widehat{f} \colon \Nomega^d \to \Nomega^d$ satisfies the
  following property: for every $\vec{x}, \vec{x}' \in \Nomega^d$ such
  that $\vec{x}$ and $f(\vec{x})$ contain the same number of $\omega$
  (in terms of ideals, $\vec{x}$ and $f(\vec{x})$ have the same
  level), the following holds:
  \[\vec{x} < \vec{x}' \implies f(\vec{x}) < f(\vec{x}').\] A
  \emph{leveled-increasing recursive net} is a finite set of
  leveled-increasing recursive partial functions.
\end{defi}

Let us remark that the composition of two leveled-increasing partial
functions is still a leveled-increasing partial function, hence the
associated transition system is a WSTS with leveled-strong-strict
monotonicity.

\begin{prop}\label{level}
  Leveled-increasing recursive nets are very-WSTS\@.
\end{prop}

\begin{proof}
  Let $N$ be a leveled-increasing recursive net. By hypothesis, $\S_N$
  has strong monotonicity since the partial functions of $N$ are
  nondecreasing. Moreover, $\S_N$ can be shown to be a deterministic
  WSTS, and $\S_{\comp{N}}$ is a deterministic WSTS with
  leveled-strong-strict monotonicity because finite composition of
  partial functions in $N$ is leveled-increasing. Finally, the set of
  states is $\N^d$ and we know that $\ideals{\N^d}$ has finitely many
  levels. Therefore, $\S_N$ is a very-WSTS\@.
\end{proof}

Since Petri nets (or vector addition systems with/without states) and
strongly increasing $\omega$-recursive nets are leveled-increasing
recursive nets by definition, to prove our claim it is sufficient to
prove that post-self-modifying nets and $\omega$-Petri nets are
leveled increasing recursive nets.

\begin{prop}\label{post}
  Post-self-modifying nets are very-WSTS\@.
\end{prop}

\begin{proof}
  Let $N$ be a post-self-modifying net. Let us prove that each partial
  function $f$ occurring on a transition $t$ of $N$ is
  leveled-increasing. Recall that $f(\vec{x}) = \vec{A} \cdot \vec{x}
  + \vec{b}$ where $\vec{A}$ is greater or equal to the identity
  matrix componentwise, and $\vec{b} \in \Z^d$.

  Recall that we want to show that for every $\vec{x}, \vec{x}' \in
  \Nomega^d$ such that $\vec{x}$ and $f(\vec{x})$ contain the same
  number of $\omega$'s, the following holds:
  \[\vec{x} < \vec{x}' \implies f(\vec{x}) < f(\vec{x}').\] Let
  $\vec{x}, \vec{x}' \in \Nomega^d$ be such that $\vec{x}$ and
  $f(\vec{x})$ contain the same number of $\omega$'s and such that
  $\vec{x} < \vec{x}'$. Since $\vec{A} \geq \vec{0}$, we have
  $f(\vec{x}) \leq f(\vec{x}')$. Moreover, there exists a least index
  $1 \leq \ell \leq d$ such that $\vec{x}(\ell) < \vec{x}'(\ell)$, and
  in particular $\vec{x}(\ell) \neq \omega$. Since $\vec{A}$ is
  greater or equal to the identity matrix, then $\vec{y}(i) = \omega$
  implies $f(\vec{y})(i) = \omega$ for every $\vec{y}$, \ie $\omega$'s
  cannot disappear. Since, by hypothesis, the number of $\omega$'s is
  the same in $\vec{x}$ and $f(\vec{x})$, this means that $f$ does not
  create new $\omega$'s in a new position and hence $\vec{x}$ and
  $f(\vec{x})$ have exactly the same $\omega$'s in the same
  positions. Thus, $\vec{x}(\ell) \neq \omega$ and $f(\vec{x})(\ell)
  \neq \omega$.

  Since $\vec{A}(\ell, \ell) \geq 1$ and $\vec{x}'(\ell) >
  \vec{x}(\ell)$, we deduce that
  \[\vec{A}(\ell, \ell) \cdot \vec{x}'(\ell) > A(\ell, \ell) \cdot
  \vec{x}(\ell) \neq \omega.\]

  Moreover, for every $1 \leq j \leq d$:
  \[A(\ell, j) \cdot \vec{x}'(j) \geq A(\ell, j) \cdot \vec{x}(j) \neq
  \omega\ \text{since $f(\vec{x})(\ell) \neq \omega$}.\] Thus,
  $f(\vec{x}')(\ell) = \sum_{j = 1}^d A(\ell, j) \cdot \vec{x}'(j) >
  f(\vec{x})(\ell)$. Therefore, we conclude that $f(\vec{x}') >
  f(\vec{x})$ and hence that $f$ is leveled-increasing. Hence, from Proposition~\ref{level}, we deduce that $N$ is a very-WSTS\@.
\end{proof}

\begin{prop}\label{omega}
  $\omega$-Petri nets are very-WSTS\@.
\end{prop}

\begin{proof}
  Let $N$ be a $\omega$-Petri net with places $P$ and transitions
  $T$. Let $t \in T$ be a transition of $N$. Let $f \colon \N^P \to
  \N^P$ be the partial function such that $f(\vec{x})$ is the marking
  obtained by firing $t$ in $\vec{x}$, provided that $t$ is enabled in
  $\vec{x}$. Let us prove that $f$ is leveled-increasing. Let
  $\vec{x}, \vec{x}' \in \Nomega^P$ be such that $\vec{x} < \vec{x}'$
  and
  \begin{align}
    f(\vec{x})\ \text{contains the same number of $\omega$'s
      as}\ \vec{x}\label{eq:same:omegas}
  \end{align}
  We need to show that $f(\vec{x}) < f(\vec{x}')$.

  Let $p \in P$. Note that $\vec{x}(p) = \omega$ implies
  $f(\vec{x})(p) = \omega$, \ie\ $t$ cannot remove an $\omega$ in
  $\comp{S}$. Let $\mathrm{Pre}(p, t)$ and $\mathrm{Post}(p, t)$
  denote respectively the number of tokens consumed and produced by
  $t$ in $p$. Let $\sim$ stand for $\leq$ or $<$ depending on whether
  $\vec{x}(p) \leq \vec{x}'(p)$ or $\vec{x}(p) < \vec{x}'(p)$. It
  suffices to show that $f(\vec{x})(p) \sim f(\vec{x}')(p)$. We make a
  case distinction on whether $\vec{x}(p)$ equals $\omega$ or
  not. \medskip

  \noindent \underline{Case ``$\vec{x}(p) \neq \omega$''}: Let
  \begin{align*}
  a & \defeq
  \begin{cases}
    \mathrm{Pre}(p, t) & \text{if}\ \mathrm{Pre}(p, t) \neq \omega, \\
    0 & \text{otherwise}.
  \end{cases} &
  b &\defeq \mathrm{Post}(p, t).
  \end{align*}
  We must have $b \neq \omega$ since we would otherwise have
  $f(\vec{x})(p) = \omega$ which would
  contradict~\eqref{eq:same:omegas}. Thus, $f(\vec{x})(p) = \vec{x}(p)
  - a + b \sim \vec{x}'(p) - a + b = f(\vec{x}')(p)$. \medskip

  \noindent \underline{Case ``$\vec{x}(p) = \omega$''}: Since
  $\vec{x}'(p) \geq \vec{x}(p) = \omega$, we have $\vec{x}'(p) =
  \omega$. Therefore, $f(\vec{x})(p) = \omega = f(\vec{x}')(p)$ and we
  are done.
Hence, from Proposition~\ref{level}, we deduce that $N$ is a very-WSTS\@.
\end{proof}

By the two previous propositions, we obtain the following:

\begin{cor}
  Petri nets (or vector addition systems with/without states),
  $\omega$-Petri nets, post-self-modifying nets and strongly
  increasing $\omega$-recursive nets are very-WSTS\@.
\end{cor}

Note that very-WSTS do not include transfer Petri nets, since
$\comp{\S}$ does not have leveled-strong-strict monotonicity, and
unordered data Petri nets, since $\ideals{X}$ has infinitely many
levels. Observe that $\comp\S$ may be deterministic (and finitely
branching) even when $\S$ is not, and even when $\S$ is not finitely
branching, as the example of $\omega$-Petri nets shows.

\begin{algorithm}[h]
  \DontPrintSemicolon%

  \KwSty{initialize} a tree $\T$ with root $r \colon I_0$\;

  \While{$\T$ contains an unmarked node $c \colon I$}{
    \If{$c$ has an ancestor $c'\colon I'$ s.t. $I' = I$}{
      \KwSty{mark} $c$ \tcc*[r]{stop exploration}
    }
    \Else{
      \If{$c$ has an ancestor $c' \colon I'$ such that $I' \subset I$}{
        \KwSty{let} $c'$ be the closest such ancestor\;
        $w \leftarrow \text{sequence of labels from $c'$ to $c$}$\;

        \If{$w^\infty(I) \neq I$}{
          \KwSty{replace} $c \colon I$ by $c \colon w^\infty(I)$ \tcc*[r]{accelerate}
        }
      }
      \For{$a \in \Sigma$}{
        \If{$\csucc{I, a} \not= \emptyset$}{
          \KwSty{add} arc labeled by $a$ from $c$ to a new child $d
          \colon a(I)$
        }
      }
      \KwSty{mark} $c$
    }
  }
  \Return{$\T$}
  \caption{Ideal Karp-Miller algorithm.}\label{alg:km:highlevel}
\end{algorithm}

We now present the \emph{Ideal Karp-Miller algorithm
(IKM)}\footnote{Note that the algorithm given here is slightly more
general and simplified than the one that appeared in the preliminary
version of this paper~\cite{BFG17}. Here we allow for some nested
accelerations while this was explicitly disallowed in the algorithm
of~\cite{BFG17}.} for very-WSTS in
Algorithm~\ref{alg:km:highlevel}. The algorithm starts from an ideal
$I_0$, successively computes its successors in $\comp{\S}$ and
performs accelerations as in the classical Karp-Miller algorithm for
Petri nets. For every node $c \colon I$ of the tree built by the
algorithm, let $\ideal{c} \defeq I$ and $\lvl{c} \defeq
\lvl{I}$. Let us first show that the algorithm terminates.

\begin{thm}
  Algorithm~\ref{alg:km:highlevel} terminates for very-WSTS\@.
\end{thm}

\begin{proof}
  Since $\ideals{X}$ has finitely many levels, $\lvl{I} \neq \infty$
  for every $I \in \ideals{X}$. Moreover,
  \begin{align}
    \lvl{c} \text{ is non-decreasing on each branch of }
    \T, \label{eq:level:nondecr}
  \end{align}
  that is: for every branch $c_0, c_1, \ldots$ of $\T$, we have
  $\lvl{c_0} \leq \lvl{c_1} \leq \cdots$. This observation follows
  from Proposition~\ref{prop:ideals:levels} combined with the fact
  that the algorithm constructs a node's ideal either from applying a
  transition to its parent's ideal, or by performing an acceleration.

  The rest of the argument is as for the classical Karp-Miller
  algorithm.  Suppose the algorithm does not terminate. Let $\T_n$ be
  the finite tree obtained after $n$ iterations. The infinite sequence
  $\T_0, \T_1, \ldots$ defines a unique infinite tree $\T_\infty =
  \bigcup_{n \in \N} \T_n$. Since $\comp{\S}$ is finitely branching,
  $\T_\infty$ is also finitely branching. Therefore, $\T_\infty$
  contains an infinite path $c_0, c_1, \ldots$ by K\"{o}nig's
  lemma. By~\eqref{eq:level:nondecr}, and since $\ideals{X}$ has
  finitely many levels, there exists $k, m \in \N$ such that
  \begin{align}
    \lvl{c_{k}} = \lvl{c_{k+1}} = \cdots = m. \label{eq:stable:level}
  \end{align}
  Since $\comp{\S}$ is a WSTS, the set $\ideals{X}$ is
  well-quasi-ordered, and hence we can find two indices $i, j$ such
  that $k \leq i < j$ and $\ideal{c_i} \subseteq \ideal{c_j}$. If
  $\ideal{c_i} = \ideal{c_j}$, then line~3 of the algorithm would have
  stopped the exploration of the path. Therefore, $\ideal{c_i} \subset
  \ideal{c_j}$. Let $\ell \in \N$ be the largest index such that $\ell
  < j$ and $\ideal{c_\ell} \subset \ideal{c_j}$. Let $w \in \Sigma^+$
  be the sequence of labels from $c_\ell$ to $c_j$. Note that $k \leq
  i \leq \ell < j$. Therefore, by~\eqref{eq:stable:level} and
  Proposition~\ref{prop:ideals:levels}, no acceleration occurred
  between $c_\ell$ and $c_j$, and hence \[\ideal{c_\ell} \ctrans{w}
  \ideal{c_j}.\] Let $I \defeq \ideal{c_\ell}$ and $J =
  \ideal{c_j}$. By~\eqref{eq:stable:level}, $\lvl{I} =
  \lvl{J}$. Therefore, by leveled-strong-strict monotonicity of
  $\comp{S}$, the sequence $J, w(J), w^2(J), \ldots$ is an
  acceleration candidate, and hence $w^\infty(J) \not= J$. Thus,
  line~10 has been executed on $c_j$, which implies that $\lvl{I} <
  \lvl{J}$ by Proposition~\ref{prop:ideals:levels}. This
  contradicts~\eqref{eq:stable:level}, which completes the proof.
\end{proof}

\subsection{Properties of the algorithm}\label{ssec:ikm:properties}

Let $\T_I$ denote the tree returned by
Algorithm~\ref{alg:km:highlevel} on input $(\S, I)$. Let $D_I \defeq
\bigcup_{c \in \T_I} \ideal{c}$. We claim that $D_I =
\downc{\ssucc[*]{I}}$. Instead of proving this claim directly, we take
traces into consideration and prove a stronger statement. We define
two word automata that will be useful for this purpose.

\begin{defi}
  The \emph{stuttering automaton}\footnote{We use the term
    \emph{stuttering} as paths of the automaton correspond to
    stuttering paths of~\cite{GHPR15}.} is the finite word automaton
  $\A_I$ obtained by making all of the states of $\T_I$ accepting, by
  taking the root $r$ as the initial state, and by taking the arcs of
  $\T_I$ as transitions, together with the following additional
  transitions: \begin{itemize} \item If a leaf $c$ of $\T_I$ has an
    ancestor $c'$ such that $\ideal{c} = \ideal{c'}$, then a
    transition from $c$ to $c'$ labeled by $\varepsilon$ is added to
    $\A_I$.  \end{itemize} The \emph{Karp-Miller automaton} is the
  automaton $\K_I$ obtained by extending $\A_I$ as
  follows: \begin{itemize} \item If a node $c$ of $\T_I$ has been
    accelerated because of an ancestor $c'$, then a transition from
    $c$ to $c'$ labeled by $\varepsilon$ is added to
    $\K_I$.  \end{itemize}
\end{defi}

\noindent
Both $\A_I$ and $\K_I$ can be computed from $\T_I$. Moreover, they
give precious information about the traces of $\S$. Let $\lang{\A_I}$
and $\lang{\K_I}$ denote the language over $\Sigma$ accepted by $\A_I$
and $\K_I$. Recall that $\preceq$ denotes the subword ordering. We
will show the following theorem:

\begin{thm}\label{thm:s:akm}
  For every very-WSTS $\S = (X, \trans{\Sigma}, \leq)$ and $I \in
  \ideals{X}$, \[D_I = \downc{\ssucc[*]{I}},\ \traces{\S}{I} \subseteq
  \lang{\A_I} \text{ and } \lang{\K_I} \subseteq
  \downc_{\preceq}{\traces{\S}{I}}.\] In particular, for every $x \in
  X$, $D_{\downc{x}} = \downc{\ssucc[*]{x}}$,
  $\downc_{\preceq}{\lang{\K_{\downc{x}}}} =
  \downc_{\preceq}{\traces{\S}{x}}$, and
  $\downc_{\preceq}{\traces{\S}{x}}$ is a computable regular language.
\end{thm}

The proof of Theorem~\ref{thm:s:akm} follows from the forthcoming
Propositions~\ref{prop:cover:a} and~\ref{prop:cover:b} describing the
relations between traces of $\A_I$ and $\K_I$ with traces of $\S$ and
$\comp{\S}$. We write $c \ttrans{w} c'$, $c \stuttrans{w} c'$ and $c
\atrans{w} c'$ whenever node $c'$ can be reached by reading $w$ from
$c$ in $\T_I$, $\A_I$ and $\K_I$ respectively.

\begin{prop}\label{prop:cover:a}
  Let $\S = (X, \trans{\Sigma}, \leq)$ be a very-WSTS and let $I_0 \in
  \ideals{X}$. For every $y, z \in X$, $w \in \Sigma^*$ and $c \in
  \A_{I_0}$, if $y \trans{w} z$ and $y \in \ideal{c}$, then there
  exists $d \in \A_{I_0}$ such that $c \stuttrans{w} d$ and $z \in
  \ideal{d}$.
\end{prop}

\begin{proof}
  The proof is by induction on $|w|$. If $|w| = 0$, then $w =
  \varepsilon$, which implies $z = y$. Thus, it suffices to take $d
  \defeq c$.

  Assume $|w| > 0$ and that the claims holds for words of length less
  than $|w|$. There exist $u \in \Sigma^*$, $a \in \Sigma$ and $y' \in
  X$ such that $w = ua$ and $y \trans{u} y' \trans{a} z$. By induction
  hypothesis, there exists a node $c' \in \A_{I_0}$ such that $c
  \stuttrans{u} c'$ and $y' \in \ideal{c'}$. Let $I \defeq
  \ideal{c'}$. Since $y' \trans{a} z$ and $y' \in I$, there exists
  some $J \in \ideals{X}$ such that $z \in J$ and $I \ctrans{a} J$. If
  $c'$ has a successor under $a$ labeled by $J$, then we are
  done. Otherwise, there are two cases to consider.

  \begin{itemize}
  \item If $c'$ has no successor under $a$, then $c'$ must be a leaf
    of $T_{I_0}$. Thus, $c'$ has an ancestor $c''$ in $T_{I_0}$ such
    that $\ideal{c'} = \ideal{c''}$. Thus, $c' \stuttrans{\varepsilon}
    c''$. Now, $c''$ has a successor $d$ under $a$, otherwise it would
    also be a leaf of $T_{I_0}$, which is impossible. Therefore, $J =
    \ideal{d}$, and hence $c \stuttrans{u} c' \stuttrans{\varepsilon}
    c'' \stuttrans{a} d$ and $z \in \ideal{d}$.\medskip

  \item If $c$ has a successor $d$ under $a$, then $J$ has been
    accelerated. Therefore, $\ideal{d} = v^\infty(J)$ for some $v \in
    \Sigma^+$. By definition of accelerations, $J \subseteq
    v^\infty(J)$. Therefore, $c \stuttrans{u} c' \stuttrans{a} d$ and
    $y \in \ideal{d}$.\qedhere
  \end{itemize}
\end{proof}

\begin{prop}\label{prop:cover:b}
  Let $\S = (X, \trans{\Sigma}, \leq)$ be a very-WSTS and let $I_0 \in
  \ideals{X}$. For every $z \in X$, $w \in \Sigma^*$ and $c, d \in
  \K_{I_0}$, if $c \atrans{w} d$ and $z \in \ideal{d}$, then there
  exist $y \in \ideal{c}$, $w' \succeq w$ and $z' \geq z$ such that $y
  \trans{w'} z'$.
\end{prop}

\begin{proof}
  The proof is by induction on $|w|$. If $|w| = 0$, then $w =
  \varepsilon$. We stress the fact that even though $w$ is empty, $d$
  might differ from $c$ since $\K_{I_0}$ contains
  $\varepsilon$-transitions. However, by definition of $\K_{I_0}$, we
  know that $\ideal{d} \subseteq \ideal{c}$. Therefore, $z \in
  \ideal{c}$, and we are done since $z \trans{\varepsilon} z$.

  Suppose that $|w| > 0$. Assume the claim holds for every word of
  length less than $|w|$. There exist $u, v \in \Sigma^*$, $a \in
  \Sigma$ and $d' \in \K_{I_0}$ such that $w = uav$, $c \atrans{u} d'
  \atrans{a} d \atrans{v} d$ and $d'$ is the parent of $d$ in
  $T_{I_0}$. Let $I \defeq \ideal{c}$, $J \defeq \ideal{d'}$, $K
  \defeq \ideal{d}$, and $K' \defeq a(J)$. By induction hypothesis,
  there exist $y_K \in K$, $v' \succeq v$ and $z' \geq z$ such that
  $y_K \trans{v'} z'$.
  \begin{itemize}
  \item If $K = K'$, then $J \ctrans{a} K$. By definition of
    $\ctrans{a}$, there exist $y_J \in J$ and $y_K' \geq y_K$ such
    that $y_J \trans{a} y_K'$. By induction hypothesis, there exist
    $y_I \in I$, $u' \succeq u$ and $y_J' \geq y_J$ such that $y_I
    \trans{u'} y_J'$. By strong monotonicity of $\S$, there exists $z''
    \geq z'$ such that $y_I \trans{u'av'} z''$. We are done since
    $u'av' \succeq uav$.\medskip

  \item If $K \not= K'$, then $K$ was obtained through an
    acceleration. Therefore, $K = \sigma^\infty(K')$ for some $\sigma
    \in \Sigma^+$. This implies that $y_K \in \sigma^k(K')$ for some
    $k \in \N$. Let $L \defeq \sigma^k(K')$. Note that $J \ctrans{a}
    K' \ctrans{\sigma^k} L$. By Proposition~\ref{prop:run:comp}(2), 
    there exist $y_J \in J$ and $y_K' \geq y_K$ such that $y_J
    \trans{a \sigma^k} y_K'$. By induction hypothesis, there exist
    $y_I \in I$, $u' \succeq u$ and $y_J' \geq y_J$ such that $y_I
    \trans{u'} y_J'$. By strong monotonicity of $\S$, there exists
    $z'' \geq z'$ such that $y_I \trans{u' a \sigma^k v'}
    z''$.\qedhere
  \end{itemize}
\end{proof}

\noindent
We may now prove the main theorem of this section:

\begin{proof}[Proof of Theorem~\ref{thm:s:akm}] \leavevmode
  \begin{enumerate}
    \item $\subseteq$:\ Let $y \in D_I$. There exist $w \in \Sigma^*$
      and $c \in \K_I$ such that $r \atrans{w} c$ and $y \in
      \ideal{c}$. By Proposition~\ref{prop:cover:b}, there exist $x
      \in I$, $w' \succeq w$ and $y' \geq y$ such that $x \trans{w'}
      y$. Hence, $y \in \ssucc[*]{x} \subseteq \ssucc[*]{I_0}
      \subseteq \downc{\ssucc[*]{I}}$.\medskip

      $\supseteq$:\ Let $y \in \downc{\ssucc[*]{I}}$. There exist $x
      \in I$, $w \in \Sigma^*$ and $y' \geq y$ such that $x \trans{w}
      y'$. By Proposition~\ref{prop:cover:a}, there exists a node $c
      \in \A_I$ such that $r \stuttrans{w} c$ and $y' \in
      \ideal{c}$. Since ideals are downward closed, $y \in \ideal{c}$
      which implies that $y \in D_I$.\medskip

    \item Let $w \in \traces{\S}{I}$. There exist $x \in I$ and $y \in
      X$ such that $x \trans{w} y$. By Proposition~\ref{prop:cover:a},
      there exists a node $c \in \A_I$ such that $r \stuttrans{w} c$
      and $y \in \ideal{c}$. Therefore, $w \in \lang{\A_I}$. \medskip

    \item Let $w \in \lang{\K_I}$. There exists a node $c \in \K_I$
      such that $r \atrans{w} c$. Let $y \in \ideal{c}$. By
      Proposition~\ref{prop:cover:b}, there exists $x \in I$, $w'
      \succeq w$ and $y' \geq y$ such that $x \trans{w'} y'$.
      Therefore, $w \in \downc_{\preceq}{\traces{\S}{I}}$ since $w
      \preceq w'$.\qedhere
  \end{enumerate}
\end{proof}

\begin{cor}\label{cor:ikm:traces}
  For every very-WSTS $\S = (X, \trans{\Sigma}, \leq)$ and every state
  $x \in X$,
  \[D_{\downc{x}} = \downc{\ssucc[*]{x}} \text{ and }
  \downc_{\preceq}{\lang{\K_{\downc{x}}}} =
  \downc_{\preceq}{\traces{\S}{x}}.\] In particular,
  $\downc_{\preceq}{\traces{\S}{x}}$ is a regular language computable
  from $\S$ and $x$.
\end{cor}

\begin{proof}\leavevmode
  \begin{itemize}
  \item By Theorem~\ref{thm:s:akm}, we have $D_{\downc{x}} =
    \downc{\ssucc[*]{\downc{x}}}$. Moreover, by strong monotonicity of
    $\S$, we have $\downc{\ssucc[*]{\downc{x}}} =
    \downc{\ssucc[*]{x}}$.\medskip

  \item By Theorem~\ref{thm:s:akm}, we have
    \[\traces{\S}{\downc{x}} \subseteq \lang{\A_{\downc{x}}} \subseteq
    \lang{\K_{\downc{x}}} \subseteq \downc_{\preceq}{\traces{\S}{\downc{x}}}.\]
    Therefore $\downc_{\preceq}{\traces{\S}{\downc{x}}} =
    \downc_{\preceq}{\lang{\A_{\downc{x}}}} =
    \downc_{\preceq}{\lang{\K_{\downc{x}}}}$. Moreover, by strong monotonicity
    of $\S$, we have $\traces{\S}{\downc{x}} =
    \traces{\S}{x}$.\qedhere
  \end{itemize}
\end{proof}

\subsection{Effectiveness of the algorithm}

The Ideal Karp-Miller algorithm can be implemented provided that
\begin{enumerate}
\item ideals can be effectively manipulated, \ie, the set of encodings
  of $\ideals{X}$ is recursive and the encoding of $\downc{x}$ is
  computable from $x \in X$ (see~\cite{BFM17} for a formal treatment
  of encodings),

\item inclusion of ideals can be tested,

\item $\csucc{I, a}$ can be computed for every ideal $I$ and $a \in
  \Sigma$, and

\item $w^\infty(I)$ can be computed for every ideal $I$ and sequence
  $w \in \Sigma^+$.
\end{enumerate}

\noindent
A class of WSTS satisfying~(1--3) is called
\emph{completion-post-effective}, and a class satisfying~(4) is called
\emph{$\infty$-completion-effective}. By Theorem~\ref{thm:s:akm}, we
obtain the following:

\begin{thm}
  Let $\C$ be a completion-post-effective and
  $\infty$-completion-effective class of very-WSTS\@. The ideal
  decomposition of $\downc{\ssucc[*]{x}}$ can be computed for every
  $\S = (X, \trans{}, \leq) \in C$ and $x \in X$. In particular,
  coverability for $\C$ is decidable.
\end{thm}


\section{Model checking liveness properties for very-WSTS}
In this section, we show how the Ideal Karp-Miller algorithm can be
used to test whether a very-WSTS violates a liveness property
specified by an LTL formula. We follow classical constructions that
have also been adapted to WSTS by Emerson and Namjoshi~\cite{EN98}
without effectiveness constraints. Testing that $\S$ violates a
property $\varphi$ amounts to constructing a Büchi automaton
$\B_{\neg \varphi}$ for $\neg \varphi$ and testing whether
$\B_{\neg \varphi}$ accepts an infinite trace of $\S$. We first show
that repeated coverability is decidable for very-WSTS under some
effectiveness hypotheses. Then, we show how LTL model checking reduces
to repeated coverability.

\subsection{Deciding repeated coverability}\label{ssec:dec:rep:cov}

Let $\S = (X, \trans{\Sigma}, \leq)$ be a WSTS, let $x \in X$ and let
$I \in \ideals{X}$. We say that $w \in \Sigma^*$ is \emph{$(I,
x)$-increasing} if there exist $y \in I$ and $z \in X$ such that
$y \trans{w} z$ and $x \leq y \leq z$. We establish a necessary and
sufficient condition for repeated coverability in terms of the
stuttering automaton and $(I, x)$-increasing sequences:

\begin{prop}\label{prop:repcov:charac}
  Let $\S = (X, \trans{\Sigma}, \leq)$ be a very-WSTS and let $x,
  y \in X$. State $y$ is repeatedly coverable from $x$ if and only if
  there exist a state $c$ of the stuttering automaton
  $\A_{\downc{x}}$, and a sequence $w \in \Sigma^+$, such that
  $c \stuttrans{w} c$ and $w$ is $(\ideal{c}, y)$-increasing.
\end{prop}

\begin{proof}
  ($\Rightarrow$) Assume $y$ is repeatedly coverable from $x$. There
  exist $y_0, y_1, \dots \in X$, $v_0 \in \Sigma^*$ and $v_1,
  v_2, \ldots \in \Sigma^+$ such that \[x \trans{v_0} y_0 \trans{v_1}
  y_1 \trans{v_2} \cdots\] and $y_i \geq y$ for every $i \in \N$. By
  Proposition~\ref{prop:cover:a}, there exist $c_0,
  c_1, \ldots \in \A_{\downc{x}}$ such that \[r \stuttrans{v_0}
  c_0 \stuttrans{v_1} c_1 \stuttrans{v_2} \cdots\] and
  $y_i \in \ideal{c_i}$ for every $i \in \N$. Since $\A_{\downc{x}}$
  is finite, there exists $c \in \A_{\downc{x}}$ such that $I
  = \{i \in \N : c_i = c\}$ is infinite. Since $X$ is
  well-quasi-ordered, there exist $i, j \in I$ such that $i < j$ and
  $y_i \leq y_j$. Let $w \defeq v_{i+1} \cdots v_j$. We have
  $c \stuttrans{w} c$ and $|w| > 0$.  Moreover, $w$ is $(\ideal{c},
  y)$-increasing since $y_i \in \ideal{c}$, $y_i \trans{w} y_j$ and
  $y \leq y_i \leq y_j$. \medskip

  ($\Leftarrow$) Let $c \in \A_{\downc{x}}$ and $w \in \Sigma^+$ be
  such that $c \stuttrans{w} c$ and $w$ is $(\ideal{c},
  y)$-increasing. Since $w$ is $(\ideal{c}, y)$-increasing, there
  exist $y' \in \ideal{c}$ and $y'' \in X$ such that $y \leq y' \leq
  y''$ and \begin{align} y' \trans{w}
  y''. \label{eq:ideal:pos} \end{align} Let $u \in \Sigma^*$ be the
  (unique) path from $r$ to $c$ in $\T_{\downc{x}}$. By
  Proposition~\ref{prop:cover:b}, there exist $x' \in \ideal{r}$,
  $u' \succeq u$ and $z \geq y'$ such that \begin{align} x' \trans{u'}
  z. \label{eq:repcov} \end{align} Since $\ideal{r} = \downc{x}$, we
  have $x' \leq x$. By~\eqref{eq:repcov} and strong monotonicity of
  $\S$, $x \trans{u'} z'$ for some $z' \geq z$. Let $y_0 \defeq
  z'$. By~\eqref{eq:ideal:pos}, $y_0 \geq y'$ and strong monotonicity
  of $\S$, we have $y_0 \trans{w} y_1$ for some $y_1 \geq y''$. Note
  that $y_1 \geq y'' \geq y'$, and hence again by strong monotonicity,
  $y_1 \trans{w} y_2$ for some $y_2 \geq y''$. By such successive
  application of strong monotonicity, we obtain $y_1, y_2,
  y_3, \ldots \in X$ such that $y_i \trans{w} y_{i+1}$ and
  $y_{i+1} \geq y''$ for every $i \in \N$. Therefore, \[x \trans{u'}
  y_0 \trans{w} y_1 \trans{w} \cdots\] and we are done since $y_0 =
  z' \geq z \geq y' \geq y$ and $y_i \geq y'' \geq y' \geq y$ for
  every $i \in \N$.
\end{proof}

Proposition~\ref{prop:repcov:charac} allows us to show the
decidability of repeated coverability under the following
effectiveness hypothesis. A class $\C$ of WSTS is
\emph{ideal-increasing-effective} if there is an algorithm that
decides the following:

\begin{center}
  \begin{tabular}{lp{10.5cm}}
    \\[0pt]

    \textsc{Input}: & $\S = (X, \trans{\Sigma}, \leq) \in \C$, $I \in
    \ideals{X}$, $x \in I$ and a finite automaton
    $A$ such that $\csucc{I}[w] \neq \emptyset$ for every $w \in
    L(A)$. \\[8pt]

    \textsc{Decide}: & does there exist $w \in L(A)$ such that $w$ is
    $(I, x)$-increasing? \\ \\
  \end{tabular}
\end{center}

Before proving decidability of repeated coverability, we first prove
two useful observations on the stuttering automaton. For every node
$c$ of an IKM tree, we define $\numaccel{c}$ as the number of
accelerations performed by Algorithm~\ref{alg:km:highlevel} from the
root $r$ of the tree to $c$ inclusively. The following holds:

\begin{prop}\label{prop:stut:numaccel}
  Let $\S = (X, \trans{\Sigma}, \leq)$ be a very-WSTS and let $I_0 \in
  \ideals{X}$. Let $c, d \in \T_{I_0}$. The following holds:
  \begin{enumerate}
    \item If $c \ttrans{*} d$ and $\ideal{c} = \ideal{d}$, then
      $\numaccel{c} = \numaccel{d}$.

    \item If $c \stuttrans{*} d$, then $\numaccel{c} \leq
      \numaccel{d}$.
  \end{enumerate}
\end{prop}

\begin{proof}\leavevmode
  \begin{enumerate} \item For the sake of contradiction, suppose that
  $\numaccel{c} \neq \numaccel{d}$. This means that at least one
    acceleration occurred between $c$ (exclusively) and $d$
    (inclusively). Let $d'$ be the first accelerated node, \ie the
    first node $d'$ for which there exists $c'$ such that
    $\numaccel{d'} = \numaccel{c'} + 1$ and \[c \ttrans{+} c'
    \ttrans{} d' \ttrans{*} d.\] By
    Proposition~\ref{prop:ideals:levels}, $\lvl{\ideal{c}} =
    \lvl{\ideal{c'}} < \lvl{\ideal{d'}} \leq \lvl{\ideal{d}}$. This is
    a contradiction since $\ideal{c} = \ideal{d}$. \bigskip

  \item Since $c$ can reach $d$, there exist a path of length $n \geq
    0$ from $c$ to $d$ in $\A_{I_0}$. Let $c_0, c_1, \dots c_n$ be the
    nodes visited by this path, where $c_0 = c$ and $c_n = d$. We
    prove the claim by induction on $n$. If $n = 0$, then $c = d$ and
    the the claim trivially holds. Assume that $n > 0$ and that the
    claims holds for paths of length $n - 1$. By induction hypothesis,
    $\numaccel{c_0} \leq \numaccel{c_{n-1}}$. If $c_n$ is an ancestor
    of $c_{n-1}$ in $\T_{I_0}$ and $\ideal{c_{n-1}} = \ideal{c_n}$,
    then we are done since $\numaccel{c_{n-1}} = \numaccel{c_n}$
    by~(1). Otherwise, $c_{n-1} \ttrans{a} c_n$ for some $a \in
    \Sigma$. By Proposition~\ref{prop:ideals:levels},
    $\numaccel{c_{n-1}} \leq \numaccel{c_n}$, and hence
    $\numaccel{c_0} \leq \numaccel{c_{n-1}} \leq
    \numaccel{c_n}$. \qedhere
  \end{enumerate}
\end{proof}

\noindent
We may now prove the decidability of repeated coverability under some
effectiveness hypotheses:

\begin{thm}\label{thm:rep:cov}
  Repeated coverability is decidable for completion-post-effective,
  $\infty$-completion-effective and ideal-increasing-effective classes
  of very-WSTS\@.
\end{thm}

\begin{proof}
  By Proposition~\ref{prop:repcov:charac}, $y$ is repeatedly coverable
  from $x$ if and only if there exist $c \in \A_{\downc{x}}$ and $w
  \in \Sigma^+$ such that
  \begin{align}
    c \stuttrans{w} c \text{ and } w \text{ is } (\ideal{c},
    y)\text{-increasing}.\label{eq:repcov:eff}
  \end{align}
  We show how~(\ref{eq:repcov:eff}) can be tested. For every $c \in
  \A_{\downc{x}}$, let $A_c$ be the automaton obtained from
  $\A_{\downc{x}}$ by taking $c$ as the initial state and the unique
  accepting state. Let $A^+_c$ be a finite automaton that recognizes
  $L(A_c) \setminus \{\varepsilon\}$. By~\eqref{eq:repcov:eff}, $y$ is
  repeatedly coverable from $x$ if and only if there exists $c \in
  \A_{\downc{x}}$ such that
  \begin{align}
    L(A^+_c) \text{ contains an } (\ideal{c}, y)\text{-increasing
      sequence}.\label{eq:contains:inc}
  \end{align}
  Let us explain how to decide~\eqref{eq:contains:inc}. First, note
  that $A^+_c$ can be constructed effectively for every $c$ using the
  fact that $\C$ is completion-post-effective and
  $\infty$-completion-effective. We may also only consider nodes $c$
  such that $y \in \ideal{c}$. Note that if $L(A^+_c)$ contains an
  $(\ideal{c}, y)$-increasing sequence, then $y \in \ideals{c}$ due to
  downward closure of ideals. Thus, when consider a node $c$, we may
  first test whether $y \in \ideal{c}$ by
  completion-post-effectiveness.

  Now, to test~\eqref{eq:contains:inc}, we may use the fact that $\C$
  is ideal-increasing-effective. In order to do so, we must show that
  for every node $c$ such that $y \in \ideal{c}$, the automaton
  $A^+_c$ is such that $\csucc{\ideal{c}}[w] \neq \emptyset$ for every
  $w \in L(A^+_c)$. Let $c$ be such that $y \in \ideal{c}$ and let $w
  \in L(A^+_c)$. We have \[c \stuttrans{w} c\] and, by
  Proposition~\ref{prop:stut:numaccel}(2), no acceleration can occur 
  along this path. Therefore, $\ideal{c} \ctrans{w} \ideal{c}$ which
  implies that $\csucc{\ideal{c}}[w] \neq \emptyset$.
\end{proof}

Let us remark that ideal-increasing-effective holds for Petri nets and
$\omega$-Petri nets, since, for these models, testing whether a finite
automaton $A$ accepts some $(I, x)$-increasing sequence amounts to
computing the Parikh image of $L(A)$, which is effectively
semilinear~\cite{Par66}:

\begin{prop}
  Petri nets (or vector addition systems with/without states) and
  $\omega$-Petri nets are ideal-increasing-effective.
\end{prop}

\begin{proof}
  Since $\omega$-Petri nets encompass all three models, we only give a
  proof for that model. For a definition of $\omega$-Petri nets, see
  either Section~\ref{sec:ikm} or~\cite{GHPR15}. Let $\S$ be an
  $\omega$-Petri net with places $P$ and transitions $T$. Let $I \in
  \ideals{\N^P}$ and $\vec{x} \in I$. Let $A$ be a finite automaton
  such that $\csucc{I}[w] \neq \emptyset$ for every $w \in L(A)$. We
  will show how to determine whether $L(A)$ contains an $(I,
  \vec{x})$-increasing sequence. Before doing so, we introduce a few
  definitions.

For every place $p$ and transition $t$ of $\S$, let $\mathrm{Pre}(p,
  t)$ and $\mathrm{Post}(p, t)$ denote respectively the number of
  tokens consumed and produced by $t$ in $p$. Let $\vec{N}$ be the
  matrix defined as follows:
  \[
  \vec{N}(p, t) \defeq
  \begin{cases}
    \mathrm{Post}(p, t) - \mathrm{Pre}(p, t)
    & \text{if}\ \mathrm{Post}(p, t) \neq
    \omega\ \text{and}\ \mathrm{Pre}(p, t) \neq \omega, \\

    \mathrm{Post}(p, t) & \text{otherwise}.
  \end{cases}
  \] Intuitively, $\vec{N}$ records the maximal increment that can be
  achieved in each place by firing transitions of $\S$. In particular,
  if $\S$ is a standard Petri net, then $\vec{N}$ is its incidence
  matrix. By abuse of notation, ``$z = y + \omega$'' with $z, y \in
  \N$ will stand for ``$z \geq y$''.

  For every word $w$, let $\parikh{w}$ be the \emph{Parikh image} of
  $w$, \ie the vector such that $\parikh{w}(a)$ is the number of
  occurrences of $a$ in $w$. Furthermore, let
  \[\parikh{A} \defeq \{\parikh{w} : w \in L(A)\},\] and let $\vec{i}
  \defeq \omega\textrm{-rep}(I)$, \ie the vector from $\Nomega^P$
  associated to $I$.

  We claim that there exists $w \in L(A)$ such that $w$ is $(I,
  \vec{x})$-increasing if and only if there exist $\vec{p} \in
  \parikh{A}$ and $\vec{y} \in \N^P$ such that
  \begin{align}
    \vec{N} \cdot \vec{p} \geq \vec{0} \text{ and } \vec{x} \leq
    \vec{y} \leq \vec{i}. \label{eq:parikh:claim}
  \end{align}

  Before proving the claim, let us see how it helps proving the
  proposition. By~\cite{Par66}, it is possible to compute from $A$ a
  Presburger-definable formula $\varphi_A$ such that
  $\varphi_A(\vec{p})$ holds if and only if $\vec{p} \in
  \parikh{A}$. Let $\varphi'(A,\vec{p},\vec{i})$, written more simply $\varphi'$, be the following Presburger-definable
  sentence\footnote{Note that Presburger arithmetic typically only
    allows for integers coefficients, while $\vec{N}$ and $\vec{i}$
    may contain $\omega$'s. However, this is not an issue since
    constraints of the form ``$\sum_{i, a_i \in \Z} a_i \cdot x_i +
    \sum_{j} \omega \cdot x_j \geq 0$'' and ``$x \leq \omega$'' can
    respectively be replaced by ``$\sum_{i, a_i \in \Z} a_i \cdot x_i
    \geq 0 \lor \bigvee_{j} x_j \geq 0$'' and ``\textit{true}''.}:
  \[\exists \vec{p} \in \N^T, \exists \vec{y} \in \N^P : \varphi_A(\vec{p}) \land
  \vec{N} \cdot \vec{p} \geq \vec{0} \land \vec{x} \leq \vec{y} \leq
  \vec{i}.\] By our claim, $\varphi'$ holds if and only if $L(A)$
  contains an $(I, \vec{x})$-increasing sequence. Thus, we derive an
  algorithm from the fact that $\varphi'$ is effective and by
  decidability of Presburger arithmetic~\cite{Pr29} (see~\cite{BM07},
  \eg, for a modern presentation in English).

  Let us now prove the claim. \medskip

  \noindent ($\Leftarrow$) Suppose there exist $\vec{p} \in \parikh{A}$
  and $\vec{y} \in \N^P$ such that~\eqref{eq:parikh:claim} holds. Let
  $w \in L(A)$ be a word such that $\vec{p} = \parikh{w}$. By
  hypothesis on $A$, $w$ is fireable from $I$ in $\comp{S}$, \ie $I
  \ctrans{w} J$ for some $J$. Let $\vec{z} \in J$. By
  Proposition~\ref{prop:run:comp}(2), there exist $\vec{y}' \in I$ and 
  $\vec{z}' \geq \vec{z}$ such that $\vec{y}' \trans{w}
  \vec{z}'$. By~\eqref{eq:parikh:claim}, we have $\vec{y} \leq
  \vec{i}$. In other words, $\vec{y} \in I$. Since $\vec{y}'$ also
  belongs to $I$, which is a directed set, there exists $\vec{y}'' \in
  I$ such that $\vec{y}'' \geq \vec{y}$ and $\vec{y}'' \geq
  \vec{y}'$. By strong monotonicity of $\S$, we have $\vec{y}''
  \trans{w} \vec{z}''$ for some $\vec{z}'' \geq \vec{z}'$. We claim
  that there exists $\vec{z}''' \geq \vec{z}''$ such that
  \begin{align}
    \vec{z}''' = \vec{y}'' + \vec{N} \cdot \vec{p}.\label{eq:incid:equal}
  \end{align}
  Vector $\vec{z}'''$ can be derived by resolving the non determinism
  of each transition $t$ occurring in the firing sequence $w$ as
  follows:
  \begin{itemize}
  \item for every place $p$ such that $\mathrm{Pre}(p, t) = \omega$,
    we make every occurrence of $t$ consume $0$ token from $p$;

  \item for every place $p$ such that $\mathrm{Post}(p, t) = \omega$,
    we make every occurrence of $t$ produce a sufficiently large
    amount of tokens in $p$, \eg\ $|\vec{z}''(p) - \vec{y}''(p)|$
    tokens.
  \end{itemize}
  This way, we have:
  \begin{align*}
    \vec{z}''' &= \vec{y}'' + \vec{N} \cdot \vec{p} &&
    \text{(by~\eqref{eq:incid:equal})} \\
    &\geq \vec{y}'' && \text{(by $\vec{N} \cdot \vec{p} \geq \vec{0}$
      from~\eqref{eq:parikh:claim}).}
  \end{align*}
  Moreover, by~\eqref{eq:parikh:claim} and by transitivity, we have
  $\vec{x} \leq \vec{y} \leq \vec{y}''$. Thus, overall, we obtain
  $\vec{y}'' \trans{w} \vec{z}'''$ and $\vec{x} \leq \vec{y}'' \leq
  \vec{z}'''$, which means that $w$ is $(I,
  \vec{x})$-increasing. \medskip

  \noindent ($\Rightarrow$) Suppose there exists $w \in L(A)$ such that
  $w$ is $(I, \vec{x})$-increasing. By definition, there exist
  $\vec{y} \in I$ and $\vec{z} \in X$ such that $\vec{y} \trans{w}
  \vec{z}$ and $\vec{x} \leq \vec{y} \leq \vec{z}$. Let us take
  $\vec{p} \defeq \parikh{w}$. By definition of $\vec{N}$, the
  following holds for $\omega$-Petri nets (and it is an equality for Petri nets):
  \begin{align}
    \vec{z} \leq \vec{y} + \vec{N} \cdot \parikh{w}.\label{eq:incid:id}
  \end{align}
  Thus, by~\eqref{eq:incid:id}, we have $\vec{y} \leq \vec{z} \leq
  \vec{y} + \vec{N} \cdot \vec{p}$. This
  implies that $\vec{N} \cdot \vec{p} \geq \vec{0}$. Moreover, the
  inequalities $\vec{x} \leq \vec{y} \leq \vec{i}$ hold by hypothesis
  and by the fact that $\vec{y} \in I$ which is equivalent to $\vec{y}
  \leq \vec{i}$.
\end{proof}

\subsection{From model checking to repeated coverability}\label{ssec:model:checking}

We conclude this section by reducing LTL model checking to repeated
coverability. Recall that a \emph{Büchi automaton} $\B$ is a
non-deterministic finite automaton $\B = (Q, \Sigma, \delta, q_0, F)$
interpreted over $\Sigma^\omega$. An infinite word is accepted by
$\B$ if it contains an infinite path from $q_0$ labeled by $w$ and
visiting $F$ infinitely often. We denote by $\lang{\B}$ the set of
infinite words accepted by $\B$.

Let $\B = (Q, \Sigma, \delta, q_0, F)$ be a Büchi automaton and let
$\S = (X, \trans{\Sigma}, \leq)$ be a WSTS\@. The product of $\B$ and
$\S$ is defined as \[\B \times \S \defeq (Q \times X, \trans{\Sigma
  \times Q}, = \times \leq)\] where $(p, x) \trans{(a, r)} (q, y)$ if
$(p, a, r) \in \delta, q = r$ and $x \trans{a} y$.  The point in
including $r$ in the label is so that the completion of $\B \times \S$
is deterministic, a requirement for very-WSTS\@. This is formalized in
the following proposition:

\begin{prop}\label{prop:product:closed}
  Let $\B = (Q, \Sigma, \delta, q_0, F)$ be a Büchi automaton and let
  $\S = (X, \trans{\Sigma}, \leq)$ be a very-WSTS\@. The product
  $\B \times \S$ is a very-WSTS\@. Moreover, it preserves
  completion-post-effectiveness, $\infty$-completion-effectiveness and
  ideal-increasing-effectiveness.
\end{prop}

\begin{proof}
  Let us show that $\B \times \S$ is a WSTS with strong
  monotonicity. Since equality is a wqo for finite sets and since wqos
  are closed under cartesian product, $= \times \leq$ is a wqo. Let
  $p, q \in Q$, $x, x', y \in X$ and $(a, r) \in \Sigma \times Q$ be
  such that
  \[ (p, x) \trans{(a, r)} (q, y) \text{ and } x' \geq y.
  \] By definition of $\B \times \S$, we have $(p, a, q) \in \delta$,
  $r = q$ and $x \trans{a} y$. By strong monotonicity of $\S$, there
  exists $y' \geq y$ such that $x' \trans{a} y'$. Therefore, \[(p,
  x') \trans{(a, r)} (q, y').\]

  It remains to show that the completion of $\B \times \S$ is a
  deterministic WSTS with leveled-strong-strict monotonicity, and that
  $\ideals{Q \times X}$ has finitely many levels. First note
  that \begin{align} \ideals{Q \times X} = \{\{q\} \times I : q \in Q,
    I \in \ideals{X}\}.\label{eq:prod:ideals} \end{align} Since
  $\ideals{X}$ has finitely many levels, it follows
  from~(\ref{eq:prod:ideals}) that $\ideals{Q \times X}$ also has
  finitely many levels. Similarly, $\ideals{Q \times X}$ is
  well-quasi-ordered by $\subseteq$ since $\ideals{X}$ is
  well-quasi-ordered by $\subseteq$ and since $Q$ is finite. We also
  note that ideal levels are preserved, i.e.\ $\lvl{\{q\} \times I} =
  \lvl{I}$ for every $q \in Q$ and $I \in \ideals{X}$.

  \medskip\noindent\emph{Leveled-strong-strict monotonicity.} Let $I,
  I', J \in \ideals{Q \times X}$, $a \in \Sigma$ and $r \in Q$ be such
  that $\lvl{I} \neq \infty$ and
  \begin{align*}
    I \subset I', I\ \ctrans{(a, r)}\ J \text{ and } \lvl{I} =
    \lvl{J}.
  \end{align*}
  By~(\ref{eq:prod:ideals}), there exist $p, q \in Q$ and $I_p, I_p',
  J_q \in \ideals{X}$ such that $I = \{p\} \times I_p$, $I' = \{p\}
  \times I_p'$ and $J = \{q\} \times J_q$. We have $I_p \subset I_p'$,
  $I_p \ctrans{a} J_q$, $q = r$ and $(p, a, r) \in \delta$. By
  leveled-strong-strict monotonicity of $\comp{\S}$, there exists
  $J_q' \in \ideals{X}$ such that $I_p' \ctrans{a} J_q'$ and $J_q
  \subset J_q'$. Let $J' \defeq \{q\} \times J_q'$. We
  obtain \[I'\ \ctrans{(a, r)}\ J' \text{ and } J \subset J'.\]

  \medskip\noindent\emph{Determinism.} Let $I, J, J' \in \ideals{Q
    \times X}$, $a \in \Sigma$ and $r \in Q$ be such that
  $I\ \ctrans{(a, r)}\ J$ and $I\ \ctrans{(a,
    r)}\ J'$. By~(\ref{eq:prod:ideals}), there exist $p, q, q' \in Q$
  and $I_p, J_q, J_{q'} \in \ideals{X}$ such that $I = \{p\} \times
  I_p$, $J = \{q\} \times J_q$ and $J' = \{q'\} \times J_{q'}$. We
  have $r = q = q'$, $I_p \ctrans{a} J_q$ and $I_p \ctrans{a}
  J_{q'}$. Since $\comp{\S}$ is deterministic, we have $J_q = J_{q'}$,
  and hence $J = J'$.

  \medskip\noindent\emph{Effectivenesses.}
  Completion-post-effectiveness is preserved due to the fact that:
  ideals can be represented by an extra finite state; testing
  $\{p\} \times I \subseteq \{q\} \times J$ simply amounts to testing
  whether $p = q$ and $I \subseteq J$; and computing
  $\csucc{\{q\} \times I, (a, r)}$ simply amounts to computing the
  successors of $q$ and $I$ under $a$ in $\B$ and $\S$
  respectively. The $\infty$-completion-effectiveness is preserved due
  to the fact that any acceleration candidate must, by definition, be
  of the form $\{q\} \times I_0, \{q\} \times I_1, \{q\} \times
  I_2, \ldots$ for some $q \in Q$, and hence that it suffices to
  perform accelerations in $\comp{S}$. Similarly,
  ideal-increasing-effectiveness is preserved because testing whether
  a sequence $w$ is $(\{p\} \times I, (q, x))$-increasing amounts to
  testing whether $w$ is $(I, x)$-increasing and whether $p = q$; this
  follows again from the fact that $Q$ is ordered by equality.
\end{proof}

For every WSTS $\S = (X, \trans{\Sigma}, \leq)$, we extend $\S$ with a
new ``minimal'' element $\bot$ smaller than every other states, \ie
\begin{align*}
  \S_\bot &\defeq (X \cup \{\bot\}, \trans{\Sigma}, \leq_\bot)
\end{align*}
where transition relations are unchanged, and $\leq_\bot \defeq\ \leq
\cup\ \{(\bot, y) : y \in X \cup \{\bot\}\}$. Adding the minimal
element $\bot$ preserves the properties of very-WSTS:\@

\begin{prop}\label{prop:s:bot}
  $\S_\bot$ is a very-WSTS for every very-WSTS $\S$. Moreover, it
  preserves completion-post-effectiveness,
  $\infty$-completion-effectiveness and
  ideal-increasing-effectiveness.
\end{prop}

\begin{proof}
  It is readily seen that $X \cup \{\bot\}$ is a wqo and that
  $\S_\bot$ preserves the strong monotonicity of $\S$. Note that
  $\ideals{X \cup \{\bot\}} = \{\bot\} \cup \{I \cup \{\bot\} : I \in
  \ideals{X}\}$. Since inclusion is a wqo for $\ideals{X}$, it is also
  a wqo for $\ideals{X \cup \{\bot\}}$. Morever, $\ideals{X \cup
    \{\bot\}}$ has as many levels as $\ideals{X}$. Let
  $\ctrans{}_{\bot}$ denote the transition relation of the completion
  of $\S_{\bot}$. For every $I, J \in \ideals{X}$ and $a \in \Sigma$,
  we have $I \ctrans{a} J$ if and only if $I \cup \{\bot\}
  \ctrans{a}_{\bot} J \cup \{\bot\}$. Therefore, the completion of
  $\S_\bot$ is also deterministic and also has leveled-strong-strict
  monotonicity. It is readily seen that effectivenesses are preserved.
\end{proof}

Taking the product of $\B$ and $\S_\bot$ allows us to test whether a
word of $\lang{\B}$ is also an infinite trace of $\S$:

\begin{prop}\label{prop:b:times:s}
  Let $\B = (Q, \Sigma, \delta, q_0, F)$ be a Büchi automaton, let $\S
  = (X, \trans{\Sigma}, \leq)$ be a very-WSTS, and let $x_0 \in
  X$. There exists $w \in \lang{\B} \cap \tracesinf{\S}{x_0}$ if and
  only if there exists $q_f \in F$ such that $(q_f, \bot)$ is
  repeatedly coverable from $(q_0, x_0)$ in $\B \times \S_\bot$.
\end{prop}

\begin{proof}
  ($\Rightarrow$) Let $w \in \lang{\B} \cap \tracesinf{\S}{x_0}$. Since
  $w \in \lang{\B}$, there exist $q_1, q_2, \ldots \in Q$ such that
  $q_0 \trans{w_1} q_1 \trans{w_2} q_2 \trans{w_3} \cdots$ and $q_i
  \in F$ for infinitely many $i \in \N$. Since $F$ is finite, there
  exists some $q_f \in F$ such that $q_i = q_f$ for infinitely many $i
  \in \N$. Since $w \in \tracesinf{\S}{x_0}$, there exist $x_1, x_2,
  \ldots \in X$ such that $x_0 \trans{w_1} x_1 \trans{w_2} x_2
  \trans{w_3} \cdots$. Therefore,
  \begin{align*}
    (q_0, x_0) \trans{(w_1, q_1)} (q_1, x_1) \trans{(w_2, q_2)} (q_2,
    x_2) \trans{(q_3, w_3)} \cdots
  \end{align*}
  which implies that $(q_f, \bot)$ is repeatedly coverable from $(q_0,
  x_0)$ in $\B \times \S_\bot$ since $x_i \geq \bot$ for every $i \in
  \N$.

  ($\Leftarrow$) Suppose $(q_f, \bot)$ is repeatedly coverable from
  $(q_0, x_0)$ in $\B \times \S_\bot$. There exist $(a_1, q_1), (a_2,
  q_2), \ldots \in \Sigma \times Q$ and $(q_1, x_1), (q_2, x_2),
  \ldots \in Q \times X$ such that
  \begin{align}
    (q_0, x_0) \trans{(a_1, q_1)} (q_1, x_1) \trans{(a_2, q_2)} (q_2,
    x_2) \trans{(a_3, q_3)} \cdots\label{eq:b:s}
  \end{align}
  and $q_i = q_f$ and $x_i \geq \bot$ for infinitely many $i \in
  \N$. By~(\ref{eq:b:s}) and by definition of $\B \times \S_\bot$, we
  have $(q_i, a_i, q_{i+1}) \in \delta \text{ and } x_i \trans{a_i}
  x_{i+1}$ for every $i \in \N$. Therefore, we conclude that $a_1 a_2
  \cdots \in \lang{\B} \cap \tracesinf{\S}{x_0}$.
\end{proof}

Theorem~\ref{thm:rep:cov} together with
Propositions~\ref{prop:product:closed},~\ref{prop:s:bot}
and~\ref{prop:b:times:s} imply the decidability of LTL model checking:

\begin{thm}\label{thm:ltl}
 LTL model checking is decidable for completion-post-effective,
 $\infty$-completion-effective and ideal-increasing-effective classes
 of very-WSTS\@.
\end{thm}

Theorem~\ref{thm:ltl} implies that LTL model checking for
$\omega$-Petri nets is decidable. This includes and strictly
generalizes decidability of termination in $\omega$-Petri
nets~\cite{GHPR15} and decidability of LTL model checking for Petri
nets.

To the best of our knowledge, we also provide the first self-contained
presentation of the decidability of LTL model checking for Petri nets
that does not rely on Rackoff techniques. The first proof for the
decidability of LTL model checking for Petri nets comes from
Esparza~\cite{Esp94}; it uses a result from Jantzen and
Valk~\cite{VJ85} on the decidability of the existence of an infinite
number of occurrences of a given transition in an infinite run. This
essentially corresponds to our general study of $(I,x)$-increasing
sequences. Moreover, to derive a $2$-EXPSPACE complexity bound,
Esparza also used the logic of Yen~\cite{Yen92} which extends Rackoff
techniques. Unfortunately, some flaws in the paper of Yen were found
later by Atig and Habermehl~\cite{AH11}. Habermehl gave the first
proof that the linear-time $\mu$-calculus is in EXPSPACE~\cite{Hab97};
his proof directly uses techniques \emph{a la Rackoff} to compute the
length of short witnesses of some infinite runs. In both previous
papers, on the decidability of LTL model checking for Petri nets, it
is not clear how the proofs found therein can be extended to general
very-WSTS\@.


\section{A characterization of acceleration levels}\label{sec:inv:levels}
In this section, we give a precise characterization of ideals that
have finitely many levels.

Let us redefine the family of sets $A_n(\ideals{X})$ introduced in
Section~\ref{ssec:levels:ideals} in a more general setting. Let $Z$ be
a well-founded partially ordered set, abstracting away from the case
$Z = \ideals X$. We say that a sequence $z_0, z_1, \ldots \in Z$ is an
\emph{acceleration candidate} if $z_1 < z_2 < \cdots$. Such an
acceleration candidate is \emph{below} $z \in Z$ if $z_i \leq z$ for
every $i \in \N$, and \emph{goes through} a set $A$ if $z_i \in A$ for
some $i \in \N$.

\begin{defi}
  Let $Z$ be a partially ordered set. Let $A_0(Z) \defeq \emptyset$. For
  every ordinal $\alpha > 0$, $A_\alpha (Z)$ is the set of elements $z
  \in Z$ such that every acceleration candidate below $z$ goes through
  $A_\beta(Z)$ for some $\beta < \alpha$.
\end{defi}

The observations made in Section~\ref{ssec:levels:ideals} still hold,
i.e.\ $A_\alpha(Z) \subseteq A_\beta(Z)$ and $A_\alpha(Z)$ is
downward-closed for every $\alpha \leq \beta$.

The \emph{rank} of $z \in Z$, denoted $\rk{z}$, is the ordinal defined
inductively by \[\rk{z} \defeq \sup\{\rk{y} + 1 : y < z\},\] where
$\sup(\emptyset) \defeq 0$. The \emph{rank} of $Z$ is defined
as \[\rk{Z} \defeq \sup\{\rk{z}+1 : z \in Z\}.\] Let us first show
that $A_n(Z)$ is exactly the set of elements of rank less than $\omega
\cdot n$. This rests on the following, which is perhaps less obvious
than it seems.

\begin{lem}\label{lemma:acc:cand}
  Let $Z$ be a countable wpo. For every $z \in Z$ such that $\rk{z}$
  is a limit ordinal, $z$ is the supremum of some acceleration
  candidate $z_0 < z_1 < \cdots$. Moreover, for any given ordinal
  $\beta < \rk{z}$, the acceleration candidate can be chosen such that
  $\beta \leq z_i$ for every $i \in \N$.
\end{lem}

\begin{proof}
  Let $\alpha \defeq \rk{z}$. A \emph{fundamental sequence} for
  $\alpha$ is a monotone sequence of ordinals strictly below $\alpha$
  whose supremum equals $\alpha$.  Fundamental sequences exist for all
  countable limit ordinals, in particular for $\alpha$, since $Z$ is
  countable (\eg see~\cite{Forster:countable:ordinals}). Pick one such
  fundamental subsequence ${(\gamma_i)}_{i \in \N}$.  Replacing
  $\gamma_i$ by $\sup (\beta, \gamma_i)$ if necessary, we may assume
  that $\beta \leq \gamma_m$ for every $i \in \N$. By the definition
  of rank, for every $i \in \N$, there is an element $z_i < z$ of rank
  at least $\gamma_i$.  Since $Z$ is well-quasi-ordered, we may
  extract a non-decreasing subsequence from ${(z_i)}_{i \in
    \N}$. Without loss of generality, assume that
  $z_0 \leq z_1 \leq \cdots$. If all but finitely many of these
  inequalities were equalities, then $z$ would be equal to $z_i$ for
  $m$ large enough, but that is impossible since $z_i < z$. We can
  therefore extract a strictly increasing subsequence from
  ${(z_i)}_{i \in \N}$. This is an acceleration sequence, its supremum
  is $z$, and $\beta \leq \gamma_i \leq z_i$ for every $i$.
\end{proof}

Note that Lemma~\ref{lemma:acc:cand} fails if $Z$ is not countable:
take $Z = \omega_1+1$, where $\omega_1$ is the first uncountable
ordinal, then $\omega_1 \in Z$ is not the supremum of countably many
ordinals $< \omega_1$. This also fails if $Z$ is not
well-quasi-ordered, even when $Z$ is well-founded: consider the set
with one root $r$ above chains of length $n$, one for each $n \in \N$:
$\rk r = \omega$, but there is no acceleration candidate below $r$.

\begin{lem}\label{lemma:rank:bound:0}
  Let $Z$ be a countable wpo, and let $n \in \N$. For every $z \in Z$,
  $\rk{z} < \omega \cdot n$ if and only if $z \in A_n(Z)$.
\end{lem}

\begin{proof}
  ($\Rightarrow$) By induction on $n$. The case $n =
  0$ is immediate. Let $n \geq 1$. Given any acceleration candidate
  $z_1 < z_2 < \cdots $ below $z$, we must have $\rk{z_1} < \rk{z_2} <
  \cdots < \rk{z}$. Since $\rk{z} < \omega \cdot n$, there exist
  $\ell, m \in \N$ with $\ell < n$ such that $\rk{z} =
  \omega.\ell+m$. Therefore, $\rk{z_i} \geq \omega \cdot \ell$ for
  only finitely many $i$. In particular, there exists some $i$ such
  that $\rk{z_i} < \omega \cdot \ell$. Since $\ell < n$, we have
  $\rk{z_i} < \omega \cdot (n-1)$. By induction hypothesis, $z_i \in
  A_{n-1}(Z)$, and hence $z \in A_n(Z)$.

  ($\Leftarrow$) We show by induction on $n$ that
  $\rk{z} \geq \omega \cdot n$ implies $z \not\in A_n(Z)$. The case
  $n = 0$ is immediate. Let $n \geq 1$. In general, $\rk{z}$ is not a
  limit ordinal, but can be written as $\alpha + \ell$ for some limit
  ordinal $\alpha$ and some $\ell \in \N$. By definition of rank, $z$
  is larger than some element of rank $\alpha + (\ell-1)$, which is
  itself larger than some element of rank $\alpha + (\ell-2)$, and so
  on. Iterating this way, we find an element $y \leq z$ of rank
  exactly $\alpha$. Since $\rk{y}$ is a limit ordinal,
  Lemma~\ref{lemma:acc:cand} entails that $y$ is the supremum of some
  acceleration candidate $z_0 < z_1 < \cdots$. Moreover, since
  $\omega \cdot (n-1) < \rk{y}$, we may assume that
  $\rk{z_i} \geq \omega \cdot (n-1)$ for every $i \in \N$. By
  induction hypothesis, $z_i \not\in A_{n-1}(Z)$ for every $i \in \N$,
  and hence $z \not\in A_n(Z)$.
\end{proof}

\begin{thm}\label{thm:rank:levels}
  Let $X$ be a countable wqo such that $\ideals{X}$ is
  well-quasi-ordered by inclusion\footnote{Recall that such a wqo is
  known as an $\omega^2$-wqo~\cite{FGL12}. That we find the ordinal
  $\omega^2$ in the statement of Theorem~\ref{thm:rank:levels} and in
  the notion of $\omega^2$-wqo seems to be coincidental.}. The
  following holds: $\ideals{X}$ has finitely many levels if and only
  if $\rk{\ideals{X}} < \omega^2$.
\end{thm}

\begin{proof}
  We apply Lemma~\ref{lemma:rank:bound:0} to $Z = \ideals X$, a wpo by
  assumption. For that, we need to show that $Z$ is countable. There
  are countably many upward-closed subsets, since they are all
  determined by their finitely many minimal elements.
  Downward-closed subsets are in one-to-one correspondence with
  upward-closed subsets, through complementation, hence are countably
  many as well, and ideals are particular downward-closed subsets.

  We conclude by noting that the following are equivalent: (1)
  $\rk{\ideals X} < \omega^2$; (2)
  $\rk{\ideals X} \leq \omega \cdot n$ for some $n \in \N$; (3)
  $A_n(\ideals{X}) = \ideals X$ for some $n \in \N$ (by
  Lemma~\ref{lemma:rank:bound:0}); (4) $\limideals{n}{X} = \emptyset$
  for some $n \in \N$.
\end{proof}

While $\rk{\ideals{\N^d}} = \omega \cdot d + 1 < \omega^2$, not all
wqos $X$ used in formal verification satisfy $\rk{\ideals X} <
\omega^2$. For example, $\rk{\ideals{\Sigma^*}} =
\omega^{|\Sigma|}+1$, for any finite alphabet $\Sigma$; a similar
result holds for multisets over $\Sigma$.

\begin{prop}\label{prop:sigma:Levels}
  $\rk{\ideals{\Sigma^*}} = \omega^{|\Sigma|}+1$ for every finite
  alphabet $\Sigma$.
\end{prop}

\begin{proof}
  Let $k \defeq |\Sigma|$. The elements of $\ideals {\Sigma^*}$ are
  \emph{word-products} $P$, defined as formal products $e_1 e_2 \cdots
  e_m$ of atomic expressions of the form $a^?$, $a \in \Sigma$, or 
  $A^*$, where $a^?$ denotes $\{a, \varepsilon\}$ and $A$ is a 
  non-empty subset of $\Sigma$~\cite{KP:age:mots,FGL-stacs2009}.
  Word-products were introduced under this name in~\cite{AbdullaCBJ04}.

  \emph{Lower bound.}  Enumerate the letters of $\Sigma$ as $a_1$,
  $a_2$, \ldots, $a_k$.  Let $A_i = \{a_1, a_2, \dots, a_i\}$.  Any
  ordinal $\alpha$ strictly less than $\omega^k$ can be written in a
  unique way as $\omega^{k-1} \cdot n_{k-1} + \omega^{k-2} \cdot
  n_{k-2} + \cdots + \omega \cdot n_1 + n_0$.  Define an ideal
  $I_\alpha$ by the word-product
  \[{(a_1^?)}^{n_0} {(a_2^?A_1^*)}^{n_1} {(a_3^?A_2^*)}^{n_2} \cdots 
  {(a_k^?A_{k-1}^*)}^{n_{k-1}}.\] The first terms, $n_0$ times $a_1^?$,  
  have a different format from the rest of the word-product. For
  uniformity of treatment, we write $a_1^?$ as $a_1^? A_0^*$ (indeed 
  $A_0^* = \emptyset^* = \{\epsilon\}$), so $I_\alpha =
  {(a_1^?A_0^*)}^{n_0} {(a_2^?A_1^*)}^{n_1} \cdots
  {(a_k^?A_{k-1}^*)}^{n_{k-1}}$.

  We claim that $\beta > \alpha$ implies $I_\beta \supset I_\alpha$.

  Let $\alpha = \omega^{k-1} \cdot n_{k-1} + \omega^{k-2} \cdot
  n_{k-2} + \cdots + \omega \cdot n_1 + n_0$ and $\beta = \omega^{k-1}
  \cdot m_{k-1} + \omega^{k-2} \cdot m_{k-2} + \cdots + \omega \cdot
  m_1 + m_0$.  The condition $\beta > \alpha$ is equivalent to the
  fact that $(m_{k-1}, m_{k-2}, \dots, m_1, m_0)$ is
  lexicographically larger than $(n_{k-1}, n_{k-2}, \dots, n_1,
  n_0)$.  Write $\beta \to \alpha$ if for some $i$ with $0\leq i < k$,
  $n_{k-1} = m_{k-1}$, $n_{k-2} = m_{k-2}$, \ldots, $n_{i+1} =
  m_{i+1}$, and $m_i = n_i+1$.  Since $>$ is the transitive closure of
  $\to$, it suffices to show that $\beta \to \alpha$ implies $I_\beta
  \supset I_\alpha$.

  Containment is proved as follows.
  $I_\beta = {(a_1^?A_0^*)}^{m_0} {(a_2^?A_1^*)}^{m_1} \cdots
  {(a_k^?A_{k-1}^*)}^{m_{k-1}}$ contains
  ${(a_{i+1}^?A_i^*)}^{m_i} {(a_{i+2}^?A_{i+1}^*)}^{m_{i+1}} \cdots
  {(a_k^?A_{k-1}^*)}^{m_{k-1}}$, because the empty word belongs to the
  removed prefix
  ${(a_1^?A_0^*)}^{m_0} {(a_2^?A_1^*)}^{m_1} \cdots
  {(a_i^?A_{i-1}^*)}^{m_{i-1}}$.  Since $m_i = n_i+1 \geq 1$, we can
  write
  ${(a_{i+1}^?A_i^*)}^{m_i} \allowbreak {(a_{i+2}^?A_{i+1}^*)}^{m_{i+1}}
  \cdots \allowbreak {(a_k^?A_{k-1}^*)}^{m_{k-1}}$ as
  $a_{i+1}^? A_i^* P$, where $P$ abbreviates
  ${(a_{i+1}^?A_i^*)}^{n_i} \allowbreak {(a_{i+2}^?A_{i+1}^*)}^{m_{i+1}}
  \allowbreak \cdots \allowbreak {(a_k^?A_{k-1}^*)}^{m_{k-1}}$.  Hence
  $I_\beta$ contains $A_i^* P$.  By the definition of $\to$, $P$ is
  equal to
  \[{(a_{i+1}^?A_i^*)}^{n_i} {(a_{i+2}^?A_{i+1}^*)}^{n_{i+1}} \cdots
  {(a_k^?A_{k-1}^*)}^{n_{k-1}}.\]  We now note that $A_i^*$ contains
  ${(a_1^?A_0^*)}^{n_0} \allowbreak {(a_2^?A_1^*)}^{n_1} \allowbreak
  {(a_3^?A_2^*)}^{n_2} \cdots \allowbreak {(a_i^?A_{i-1}^*)}^{n_{i-1}}$,
  because every word in the latter contains only letters from
  $\{a_1, a_2, \dots, a_i\} = A_i$.  Hence $A_i^* P$ contains
  ${(a_1^?A_0^*)}^{n_0} {(a_2^?A_1^*)}^{n_1} {(a_3^?A_2^*)}^{n_2} \cdots
  {(a_i^?A_{i-1}^*)}^{n_{i-1}} P$, which is equal to $I_\alpha$.  Since
  $I_\beta$ contains $A_i^* P$, we conclude.

  We now show that containment is strict.  Let $w$ be the word
  $a_1^{m_0} {(a_2a_1)}^{m_1} {(a_3a_2)}^{m_2} \cdots \allowbreak
  {(a_k a_{k-1})}^{m_{k-1}}$.  Clearly, $w$ is in $I_\beta$. To show that
  $w$ is not in $I_\alpha$, we show that
  $u {(a_{i+1}a_i)}^{n_i+1} \allowbreak {(a_{i+2}a_{i+1})}^{m_{i+1}}
  \allowbreak \cdots \allowbreak {(a_j a_{j-1})}^{m_{j-1}}$ is not in
  $L {(a_{i+1}^?A_i^*)}^{n_i}\allowbreak {(a_{i+2}^?A_{i+1}^*)}^{m_{i+1}}
  \allowbreak \cdots \allowbreak {(a_j^?A_{j-1}^*)}^{m_{j-1}}$ for any
  $j$, $i+1 \leq j \leq k$, where $u$ is an arbitrary word in $A_i^*$
  and $L$ is an arbitrary language included in $A_i^*$.  We will
  obtain $w \not\in I_\alpha$ by letting $j=k$,
  $u = a_1^{m_0} {(a_2a_1)}^{m_1} \cdots \allowbreak
  {(a_i a_{i-1})}^{m_{i-1}}$ and
  $L = {(a_1^?A_0^*)}^{n_0} {(a_2^?A_1^*)}^{n_1} {(a_3^?A_2^*)}^{n_2} \cdots
  {(a_i^?A_{i-1}^*)}^{n_{i-1}}$.  This is by induction on $j-(i+1)$.  If
  $j=i+1$, we must show that $u {(a_{i+1}a_i)}^{n_i+1}$ is not in
  $L {(a_{i+1}^?A_i^*)}^{n_i}$, and that is obvious since any word in
  $L {(a_{i+1}^?A_i^*)}^{n_i}$ can contain at most $n_i$ occurrences of
  $a_{i+1}$.  In the induction case, let
  $v = u {(a_{i+1}a_i)}^{n_i+1} \allowbreak {(a_{i+2}a_{i+1})}^{m_{i+1}}
  \allowbreak \cdots \allowbreak {(a_j a_{j-1})}^{m_{j-1}}$,
  $A = L {(a_{i+1}^?A_i^*)}^{n_i}\allowbreak
  {(a_{i+2}^?A_{i+1}^*)}^{m_{i+1}} \allowbreak \cdots \allowbreak
  {(a_j^?A_{j-1}^*)}^{m_{j-1}}$, and let us show that
  $v {(a_{j+1} a_j)}^{m_j} \not \in A {(a_{j+1}^? A_j^*)}^{m_j}$, knowing
  that $v \not\in A$ by induction hypothesis.  If
  $v {(a_{j+1} a_j)}^{m_j}$ were in $A {(a_{j+1}^? A_j^*)}^{m_j}$, there
  would be two words $v_1 \in A$ and $v_2 \in {(a_{j+1}^? A_j^*)}^{m_j}$
  such that $v {(a_{j+1} a_j)}^{m_j} = v_1 v_2$.  Since $v_2$ is a
  suffix of $v {(a_{j+1} a_j)}^{m_j}$ and is in
  ${(a_{j+1}^? A_j^*)}^{m_j}$, $v_2$ must in fact be a suffix of
  ${(a_{j+1} a_j)}^{m_j}$.  Hence $v_1$ contains $v$ as prefix.
  However, $v_1$ is in $A$ and $A$ is downward-closed, and that
  implies $v \in A$ in particular: contradiction.

  This ends our proof that $\beta > \alpha$ implies
  $I_\beta \supset I_\alpha$.  Since $I_\beta \supset I_\alpha$
  implies $\rk I_\beta > \rk I_\alpha$, an easy ordinal induction
  shows that $\rk I_\alpha \geq \alpha$ for every ordinal
  $\alpha < \omega^k$.  There is a further ideal $A_k^* = \Sigma^*$ in
  $\Sigma^*$.  It contains every $I_\alpha$, and strictly so since the
  number of occurrences of $a_k$ in any word of $I_\alpha$ is bounded
  from above by $n_{k-1}$ (where
  $\alpha = \omega^{k-1} \cdot n_{k-1} + \omega^{k-2} \cdot n_{k-2} +
  \cdots + \omega \cdot n_1 + n_0$), but there are words with
  arbitrarily many occurrences of $a_k$ in $A_k^*$.  It follows that
  the rank of $A_k^*$ in $\ideals {\Sigma^*}$ is at least
  $\sup \{\alpha+1 \mid \alpha < \omega^k\} = \omega^k$, and therefore
  that the rank of $\ideals {\Sigma^*}$ is at least $\omega^k+1$.

  \emph{Upper bound.}  Order atomic expressions by: $A^* \sqsubset
  B^*$ if and only if $A \subset B$, $a^? \sqsubset B^*$ if and only
  if $a \in B$, and no other strict inequality holds. The relation
  $\sqsubset$ is simply strict inclusion of the corresponding ideals.
  A word-product $P = e_1 e_2 \cdots e_m$ is \emph{reduced} if and
  only if the ideal $e_i e_{i+1}$ is included neither in $e_i$ nor in
  $e_{i+1}$, for every $i$, $1\leq i < m$.  Reduced word-products are
  normal forms for word-products~\cite{AbdullaCBJ04}.  On reduced
  word-products, we define two binary relations $\sqsubset^\w$ and
  $\sqsubseteq^\w$ by the following rules, and the specification that
  $\sqsubseteq^\w$ is the reflexive closure of $\sqsubset^\w$:
  \[
    \begin{prooftree}
      e P \sqsubseteq^\w  P'
      \justifies
      e P \sqsubset^\w e' P'
    \end{prooftree}
    \quad
    \begin{prooftree}
      P \sqsubset^\w P'
      \justifies
      a^? P \sqsubset^\w a^? P' 
    \end{prooftree}
    \quad
    \begin{prooftree}
      \forall i \cdot e_i \sqsubset {A'}^* \quad
      P \sqsubseteq^\w P' 
      \justifies
      e_1 \ldots e_k P \sqsubset^\w {A'}^* P'
    \end{prooftree}
    \quad
    \begin{prooftree}
      P \sqsubset^\w P' 
      \justifies
      A^* P \sqsubset^\w A^* P' 
    \end{prooftree}
  \]
  Those rules are taken from~\cite[Figure~1]{JGL-mfcs13}, and
  specialized to the case where all letters from $\Sigma$ are
  incomparable.  (That means that the rule called $(\w2)$ there never
  applies, and we have kept the remaining rules $(\w1)$,
  $(\w3)$--$(\w5)$.)  For reduced word-products $P$ and $P'$,
  $P \sqsubset^\w P'$ if and only if $P$, as an ideal, is strictly
  contained in $P'$ (loc.cit.; alternatively, this is an easy exercise
  from the characterization of [non-strict] inclusion in~\cite{AbdullaCBJ04}.)  It follows that if $P$ is strictly below $P'$
  in $\ideals {\Sigma^*}$, then $\mu (P)$ is strictly below $\mu (Q)$
  in the multiset extension of $\sqsubset$, where, for
  $P = e_1 e_2 \cdots e_m$, $\mu (P)$ is the multiset
  $\{e_1, e_2, \dots, e_m\}$, a fact already used in~\cite{JGL-mfcs13}.

  The set of atomic expressions consists of the following elements:
  elements of the form $a^?$ are at the bottom, and have rank $1$; 
  just above, we find ${\{a\}}^*$, of rank $2$, then ${\{a, b\}}^*$ of
  rank $3$, etc..  In other words, $A^*$ has rank one plus the
  cardinality of $A$.  In particular, all atomic expressions except
  $\Sigma^*$ have rank at most $k$.

  Among reduced word-products $P$, those that are different from
  $\Sigma^*$ must be of the form $e_1 e_2 \cdots e_m$ where no $e_i$
  is equal to $\Sigma^*$.  This is by definition of reduction.  Hence
  the suborder of those reduced word-products $P \neq \Sigma^*$ has
  rank less than or equal to the set of multisets
  $\{e_1, e_2, \dots, e_m\}$ where each $e_i$ has rank at most $k$
  (in the set of atomic expressions different from $\Sigma^*$).

  The rank of the set of multisets of elements, where each element has
  rank at most $k$, is at most $\omega^k$.  This is well-known, but
  here is a short argument.  We can map any multiset
  $\{e_1, e_2, \dots, e_m\}$ to the ordinal
  $\omega^{k-1} \cdot n_{k-1} + \omega^{k-2} \cdot n_{k-2} + \cdots +
  \omega \cdot n_1 + n_0$ where $n_i$ counts the number of elements
  $e_j$ of rank $i$, and we observe that this mapping is strictly
  monotone.

  It follows that the suborder of those reduced word-products $P$ that
  are different from $\Sigma^*$ has rank at most $\omega^k$.
  $\ideals {\Sigma^*}$ contains just one additional element,
  $\Sigma^*$, which is therefore of rank at most $\omega^k$.  Hence
  $\ideals {\Sigma^*}$ has rank at most $\omega^k+1$.
\end{proof}


\section{Discussion and further work}
We have presented the framework of very-WSTS, for which we have given
a Karp-Miller algorithm. This allowed us to show that ideal
decompositions of coverability sets of very-WSTS are computable, and
that LTL model checking is decidable under some additional
assumptions. We have also characterized acceleration levels in terms
of ordinal ranks. Finally, we have shown that downward traces
inclusion is decidable for very-WSTS\@.

As future work, we propose to study well-structured models beyond
very-WSTS for which there exist Karp-Miller algorithms, \eg\ unordered
data Petri nets (UDPN)~\cite{HMM14,HLLLST16}, or for which
reachability is decidable, \eg recursive Petri nets\footnote{Recursive
Petri nets are WSTS for the tree
embedding.}~\cite{DBLP:journals/acta/HaddadP07} with strict
monotonicity. It is conceivable that LTL model checking is decidable
for such models. Our approach will have to be extended to tackle this
problem.

For example, UDPN do not have finitely many acceleration
levels. To circumvent this issue, Hofman~\etal~\cite{HLLLST16} make
use of two types of accelerations that can be nested. One type is
prioritized to ensure that acceleration levels along a branch grow
``fast enough'' for the algorithm to terminate. A possible way to
apply the theory of very-WSTS to such models that are not very-WSTS
simply because they have an ideal rank larger or equal to $\omega^2$
could be to find an abstraction of the set of ideals that reduces
their rank to an ordinal strictly smaller that $\omega^2$ while
preserving suitable acceleration properties.

Moreover, observe that the IKM algorithm still terminates if, for each
branch $B = (c_0 \colon I_0, c_1 \colon I_1, \ldots)$ of the Ideal
Karp-Miller tree, the following set has rank less than $\omega^2$:
\begin{align*}
  [B] &\defeq \{I \in \ideals{X} : \exists j, k \in \N, j \leq k
  \text{ and }I_j \subseteq I \subseteq I_k\}.
\end{align*}
Indeed, the IKM algorithm terminates if and only if each branch $B$ is
finite, and the states involved in computing the branch, as well as
all needed accelerations, are all included in $[B]$. Therefore,
relaxing ``$\rk {\ideals X} < \omega^2$'' to the more technical
condition ``$\rk {[B]} < \omega^2$'' may allow one to extend the
notion of very-WSTS\@.

We know from~\cite{BFP12} that model checking of the \emph{eventually
  increasing Presburger CTL} fragment of CTL, which has been defined
by Atig and Habermehl in~\cite{AH11}, is undecidable for
post-self-modifying nets, while it is decidable for Petri
nets~\cite{AH11}. However, to the best of our knowledge, the
(un)decidability of LTL model checking for post-self-modifying nets is 
still open. One could hope to show decidability using our framework by
proving ideal-increasing-effectiveness.

It also remains to establish the computational complexity of LTL model
checking for $\omega$-Petri nets, which cannot directly be done in our
framework. It might be possible to adapt extended Rackoff techniques
as done for termination in $\omega$-Petri nets~\cite{GHPR15}.


\bibliographystyle{alpha}
\bibliography{references}

\clearpage
\section{Appendix}
\begin{proof}[Proof of Proposition~\ref{prop:run:comp}(1--3)] \leavevmode 
  \begin{enumerate}
  \item By induction on $|w|$.  When $|w|=0$, the claim is obvious.
    Otherwise, write $w$ as $av$ where $a \in \Sigma$,
    $v \in \Sigma^*$, $|v| < |w|$, and let $x \trans a z \trans v y$,
    for some state $z$.  Certainly $z$ is in $\ssucc{I}[a]$, hence in
    $\downc (\ssucc{I}[a])$.  Write the ideal decomposition of the
    latter as $\{I_1, I_2, \dots, I_n\}$.  For some $k$,
    $1\leq k\leq n$, $z$ is in $I_k$, and by definition
    $I \ctrans a I_k$.  By induction hypothesis, $I_k \ctrans v J$ for
    some ideal containing $y$, whence the result.\bigskip

  \item By induction of $|w|$ again. The case $|w| = 0$ is obvious,
    too. Otherwise, write $w$ as $av$, where $a \in \Sigma$, $v \in
    \Sigma^*$, $|v| < |w|$.  There is an ideal $K$ such that $I
    \ctrans a K \ctrans v J$, and the induction hypothesis gives us
    elements $z \in K$ and $y' \in J$, and a word $v' \in \Sigma^*$
    such that $z \trans {v'} y'$ and $y' \geq y$.  (Moreover, if $\S$
    has strong monotonicity, then $v'=v$.)  By definition of $\ctrans
    a$, $K$ is included in $\downc {\ssucc{I}[a]}$, so there are
    elements $x \in I$ and $z' \in K$ with $z' \geq z$ such that $x
    \trans a z'$.  Since $\S$ is monotonic, there is a further element
    $y'' \geq y'$ and a further word $v''$ such that $z' \trans{v''}
    y''$.  (If $\S$ is strongly monotonic, $v''=v'$, so $v''=v$.)
    This entails that $x \trans {av''} y'' \geq y$, and if $\S$ is
    strongly monotonic, $av'' = av = w$.\bigskip

  \item Let $J \in \csucc{I, w}$ and let $y \in J$. By~(2), there
    exist $x \in I$ and $y' \in X$ such that $x \trans{w} y'$ and $y'
    \geq y$. Thus, $y \in \downc{\ssucc{x, w}} \subseteq
    \downc{\ssucc{I, w}}$.  Conversely, let $y \in \downc{\ssucc{I,
        w}}$. There exist $x \in I$ and $y' \in X$ such that $x
    \trans{w} y'$ and $y' \geq y$. By~(1), there exists an ideal $J
    \supseteq \downc{y'} \supseteq \downc y$ such that $I \ctrans{w}
    J$. Thus, $J \in \csucc{I, w}$ and $y \in J$.\qedhere
  \end{enumerate}
\end{proof}

\end{document}